\documentclass[graybox,natbib,nospthms]{svmult}

\usepackage[T1]{fontenc}
\usepackage{microtype}
\usepackage{lmodern}
\usepackage{amssymb}
\usepackage{amsmath}
\usepackage{amsthm}

\usepackage{graphicx}        
\usepackage{multicol}        
\usepackage[bottom]{footmisc}

\theoremstyle{definition}
\newtheorem*{GC}{General Covariance}
\newtheorem*{DI1}{Diffeomorphism Invariance (version 1)}
\newtheorem*{DI2}{Diffeomorphism Invariance (final version)}
\newtheorem*{BIV1}{Background Independence (version 1)}
\newtheorem*{BIV2}{Background Independence (version 2)}
\newtheorem*{BIV3}{Background Independence (version 3)}

\usepackage{color}
\definecolor{darkblue}{rgb}{0,0.1,0.5}

\usepackage{url}
\usepackage{nicefrac}
\usepackage{hyperref}
\hypersetup{colorlinks,
            linkcolor=darkblue,
            anchorcolor=darkblue,
            citecolor=darkblue,
            urlcolor=darkblue}

\begin{document}

\title*{Background Independence, Diffeomorphism Invariance, and the Meaning of Coordinates\thanks{10 December 2013; revised 10 June, 2015. To appear in Lehmkuhl (ed), \emph{Towards a Theory of Spacetime Theories}, forthcoming in the Einstein Studies Series (Boston: Birk\"{a}user). Please cite the published version.}}
\titlerunning{Background Independence} 
\author{Oliver Pooley}
\authorrunning{O.\ Pooley} 
\institute{Oliver Pooley \at Oriel College, Oxford, OX1 4EW, UK.\\ \email{oliver.pooley@philosophy.ox.ac.uk}}

\maketitle

%
\abstract{Diffeomorphism invariance is sometimes taken to be a criterion of background independence.  This claim is commonly accompanied by a second, that the genuine physical magnitudes (the ``observables'') of background-independent theories and those of background-dependent (non-diffeomorphism-invariant) theories are essentially different in nature.  I argue against both claims.  Background-dependent theories can be formulated in a diffeomorphism-invariant manner.  This suggests that the nature of the physical magnitudes of relevantly analogous theories (one background free, the other background dependent) is essentially the same.  The temptation to think otherwise stems from a misunderstanding of the meaning of spacetime coordinates in background-dependent theories.}

\section*{Contents}
\setcounter{minitocdepth}{1}
\dominitoc

\section{What's so Special about General Relativity?}\label{s:special}

According to a familiar and plausible view, 
 the core of Einstein's general theory of relativity (GR) is what was, in 1915, a radically new way of understanding gravitation.  In pre-relativistic theories, whether Newtonian or specially relativistic, the structure of spacetime is taken to be fixed, varying neither in time nor from solution to solution.  Gravitational phenomena are assumed to be the result of the action of gravitational forces, diverting gravitating bodies from the natural motions defined by this fixed spacetime structure.  According to GR, in contrast, freely-falling bodies are force free; their trajectories are natural motions.  Gravity is understood in terms of a mutable spacetime structure.  Bodies act gravitationally on one another by affecting the curvature of spacetime.  ``\emph{Space acts on matter, telling it how to move.  In turn, matter reacts back on space, telling it how to curve}'' \citep[5]{mtw73}.  Note that the first of the claims in the quotation is as true in pre-relativistic theories as it is in GR, at least according to the substantivalist view, which takes spacetime structure in such a theory to be an independent element of reality.  The novelty of GR lies in the second claim: spacetime curvature varies, in time (and space) and across models, and the material content of spacetime affects how it does so.

This sketch of the basic character of GR has two, separable elements.  One is the interpretation of the metric field, $g_{ab}$, as intrinsically geometrical: gravitational phenomena are to be understood in terms of the curvature \emph{of spacetime}.  The second is the stress on the dynamical nature of the metric field: the fact that it has its own degrees of freedom and, in particular, that their evolution is affected by matter.  While I believe that both of these are genuine (and novel) features of GR, my focus in this paper is on the second.  Those who reject the emphasis on geometry are likely to claim that the second element by itself encapsulates the true conceptual revolution ushered in by GR.  Non-dynamical fields, such as the spacetime structures of pre-relativistic physics, are now standardly labelled \emph{background fields} (although which of their features qualifies them for this status is a subtle business, to be explored in what follows).  On the view being considered, the essential novelty of GR is that such background structures have been excised from physics; GR is the prototypical \emph{background-independent} theory\footnote{In what follows I focus specifically on the notion of background independence that is connected to the idea that background structures are non-dynamical fields.  In doing so, I am ignoring several other (not always closely related) definitions of background independence, including those given by \citet{gryb10} (which arises more naturally in the context of Barbour's $3$-space approach to dynamics) and by \citet{rozali09} (which arises naturally in string theory).  A more serious omission is lack of discussion of the definition given by \citet{belotbi}, which is motivated by ideas closely related to the themes of this paper.  I hope to explore these connections on another occasion.} (as it happens, a prototype yet to be improved upon).

Although this paper is about this notion of background independence, the question of the geometrical status of the metric field cannot be avoided entirely.  In arguing against the interpretation of GR as fundamentally about spacetime geometry, Anderson writes:
\begin{quote}
What was not clear in the beginning but by now has been recognised is that one does not need the ``geometrical'' hypotheses of the theory, namely, the identification of a metric with the gravitational field, the assumption of geodesic motion, and the assumption that ``ideal'' clocks measure proper time as determined by this metric. Indeed, we know that both of these latter assumptions follow as approximate results directly from the field equations of the theory without further assumptions. \cite[528]{anderson96}
\end{quote}
There is at least the suggestion here that GR differs from pre-relativistic theories not only in lacking non-dynamical, background structures but also in terms of how one of its structures, the ``gravitational field'', acquires geometrical meaning: the appropriate behaviour of test bodies and clocks can be derived, approximately, in the theory.  Does this feature of GR really distinguish it from special relativity (SR)?

Consider, in particular, a clock's property of measuring the proper time along its trajectory.  In a footnote, Anderson goes on to explain  that ``the behaviour of model clocks and what time they measure can be deduced from the equations of sources of the gravitational and electromagnetic fields which in turn follow from the field equations'' \citep[529]{anderson96}.  But the generally relativistic ``equations of sources of the gravitational and electromagnetic fields'' are, on the assumption of minimal coupling, exactly the same as the equations of motion of an analogue specially relativistic theory.\footnote{That it is only in the GR context that material fields merit the label ``sources of the gravitational field'' is, of course, irrelevant.}  It follows that whatever explanatory modelling one can perform in GR, by appeal to such equations, to show that some particular material system acts as a good clock and discloses proper time, is equally an explanation of the behaviour of the same type of clock in the context of SR.  Put differently, it is as true in SR as it is in GR that the ``geometrical'' hypothesis linking the behaviour of ideal clocks to the (in this context) non-dynamical background ``metric'' field is in principle dispensable.\footnote{In this context it is interesting to consider Fletcher's proof that the clock hypothesis holds up to arbitrary accuracy for sufficiently small light clocks \citep{fletcherLC}. As is explicit in Fletcher's paper, his result is as applicable to accelerating clocks in SR as it is to arbitrarily moving clocks in GR. Fletcher's proof assumes only that light travels on null geodesics; it does not make any assumptions about the fundamental physics, or even (specific) assumptions about the deformation of the spatial dimensions of the clock. All of this is consistent with one of the morals of the ``dynamical approach to special relativity'', defended in \citet{brown05} and \citet{brownpooley06}, that it is no more of a brute fact in SR than in GR that real rods and clocks, which are more or less complex solutions of the laws governing their constituents, map out geometrical properties in the way that they do. What Fletcher's proof illustrates is that some interesting results are nonetheless obtainable from minimalist, high-level physical assumptions. (Note that, in contrast to the position taken in \citet{brownpooley06}, I am here assuming that the structure encoded by the flat metric field of special relativity corresponds to a primitive element of reality, as was entertained in \citet[82, fn~22]{brownpooley06}.)}


\section{Einstein on General Covariance}\label{s:einstein}

The previous section's positive characterisation of GR's essential difference from its predecessors goes hand-in-hand with a negative claim:  GR does not differ from its predecessors in virtue of being a \emph{generally covariant} theory.  In particular, the general covariance of GR does not embody a ``general principle of relativity'' (asserting, for example, the physical equivalence of observers in arbitrary states of relative motion).  In contrast, the restricted, Lorentz covariance of standard formulations of specially relativistic physics \emph{does} embody the (standard) relativity principle.  In Michael Friedman's words, ``the principle of general covariance has no physical content whatever: it specifies no particular physical theory; rather it merely expresses our commitment to a certain style of formulating physical theories'' \citep[55]{friedman83}.

Notoriously, of course, Einstein thought otherwise, at least initially.\footnote{The evolution of Einstein's views is covered in detail by \citet[\S3]{norton93}.  In this section I largely follow Norton's narrative.} The restricted relativity principle of SR and Galilean-covariant Newtonian theories is the claim that the members of a special class of frames of reference, each in uniform translatory motion relative to the others, are  physically equivalent.  In such theories, although no empirical meaning can be given to the idea of absolute rest, there is a fundamental distinction between accelerated and unaccelerated motion.  Einstein thought this was problematic, and offered a thought experiment to indicate why.

Consider two fluid bodies, separated by a vast distance, rotating relative to one another about the line joining their centres.  Such relative motion is in principle observable, and so far our description of the set up is symmetric with respect to the two bodies.  Now, however, imagine that one body is perfectly spherical while the other is oblate.  A theory satisfying only the restricted principle of relativity is compatible with this kind of situation.  In such a theory, the second body might be flattened along the line joining the two bodies only because that body is rotating, not just with respect to other observable bodies, but with respect to the theory's privileged, non-accelerating frames of reference.  Einstein deemed this an inadequate explanation.  He claimed that appeal to the body's motion with respect to the invisible inertial frames was an appeal to a ``merely factitious cause.''  In Einstein's view, a truly satisfactory explanation should cite ``\emph{observable facts of experience}'' \cite[113]{einstein16}.  A theory which in turn explains the (local) inertial frames in terms of the configuration of (observable) distant masses---that is, a theory satisfying (a version of) \emph{Mach's Principle}---would meet such a requirement.

In his quest for a relativistic theory of gravity, Einstein did not attempt to implement (this version of) Mach's principle directly.  Instead he believed that the \emph{equivalence principle} (as he understood it) was the key to extending the relativity principle to cover frames uniformly accelerating with respect to the inertial frames.  In standard SR, force-free bodies that move uniformly in an inertial frame $F$ are equally accelerated by inertial ``pseudo forces'' relative to a frame $F'$ that is uniformly accelerating relative to $F$.  According to Einstein's equivalence principle, the physics of frame $F'$ is strictly identical to that of a ``real'' inertial frame in which there is a uniform gravitational field.  In other words, the same laws of physics hold in two frames that accelerate with respect to each other.  According to one frame, there is a gravitational field; according to the other, there is not.  The laws that hold with respect to both frames, therefore, must cover gravitational physics.  Einstein took it to follow that there is no fact of the matter about whether a body is moving uniformly or whether it is accelerating under the influence of gravitation.  The existence of a gravitational field becomes frame-relative, in a manner allegedly analogous to the frame-relativity of particular electric and magnetic fields in special relativity.\footnote{For a recent, sympathetic discussion of this aspect of Einstein's understanding of the equivalence principle, see \citet{janssenTWINS}.}

The equivalence principle, then, led Einstein to believe both that relativistic laws covering gravitational phenomena would extend the relativity principle and that the gravitational field would depend, in a frame-relative manner, on the metric field, $g_{ab}$.  A theory implementing a general principle of relativity would affirm the physical equivalence of frames of reference in arbitrary relative motion.  Einstein took the physical equivalence of two frames to be captured by the fact that the equations expressing the laws of physics take the same form with respect to each of them.\footnote{Recall Einstein's 1905 statement of the restricted principle of relativity: ``The laws by which the states of physical systems undergo change are not affected, whether these changes of state be referred to the one or the other of two systems of co-ordinates in uniform translatory motion'' \cite[41]{einstein05sr}.}  But general covariance is the property that a theory possesses if its equations retain their form under smooth but otherwise arbitrary coordinate transformation.  Einstein noted that such coordinate transformations strictly include ``those which correspond to all relative motions of three-dimensional systems of co-ordinates'' \cite[117]{einstein16}. He therefore maintained that any generally covariant theory satisfies a general postulate of relativity.\footnote{``Es ist klar, da\ss{} eine Physik, welche diesem Postulat [i.e., general covariance] gen\"{u}gt, dem allgemeinen Relativit\"{a}tspostulat gerecht wird'' \cite[776]{einstein16}.}

Einstein soon modified his view.  Essentially the view expressed by Friedman in the quotation given above---that any theory can be given a generally covariant formulation---was put to Einstein by  \citet{kretschmann17}.\footnote{Kretschmann's position is more subtle than the headline lesson that is standardly taken from it.  In particular, he relied on a key premise, closely analogous to the central premise of Einstein's `point-coincidence' response to his own hole argument, that the factual content of a theory is exhausted by spatiotemporal coincidences between the objects and processes it posits; see \citet[\S5.1]{norton93}.  The assumption that the basic objects of a theory must be well defined in the sense of differential geometry has come to play a similar role in modern renditions of Kretschmann's claim.}  
In his response, Einstein conceded the basic point \citep{einstein18}.  He identified three principles as at the heart of GR: (a) the (general) principle of relativity; (b) the equivalence principle; and (c) Mach's principle.  The relativity principle, at least as characterised in his reply to Kretschmann, was no longer conceived of in terms of the physical equivalence of frames of reference in various types of relative motion.  Instead it had simply become the claim that the laws of nature are  statements only about spatiotemporal coincidences, from which it was alleged to be an immediate corollary that such laws ``find their natural expression'' in generally covariant equations.  Mach's principle was also given a GR-specific rendition:  the claim was that the metric was completely determined by the masses of bodies.  

In another couple of years, as a result of findings by de~Sitter and Klein, Einstein was also forced to accept that his theory did not vindicate Mach's ideas about the origin of inertia.  His official objection to the spacetime structures of Newtonian and specially relativistic theories changed accordingly, in order to fit this new reality.\footnote{For more on the evolution of this aspect of Einstein's thinking, see \citet{brownlehmkuhlAR}.}  Einstein conceded that taking Newtonian physics at face value involves taking Newton's Absolute Space to be ``some kind of physical reality'' \citep[15]{einstein24}.  That it has to be conceived of as something real is, he says, ``a fact that physicists have only come to understand in recent years'' \citep[16]{einstein24}.  It is absolute, however, not merely in the substantivalist sense that it exists absolutely.  Now Einstein placed emphasis on the fact that it is not \emph{influenced} ``either by the configuration of matter, or by anything else'' \citep[15]{einstein24}.  This violation of the \emph{action--reaction principle}, rather than its status as an unobservable causal agent, came to be seen as what is objectionable about pre-relativistic spacetime.  In Einstein's words, ``it is contrary to the mode of thinking in science to conceive of a thing (the space-time continuum) which acts itself, but which cannot be acted upon'' \citep[62]{einstein22}.\footnote{Similarly, Anderson writes that violation of what he calls a general principle of reciprocity ``seems to be fundamentally unreasonable and unsatisfactory'' \citep[192]{anderson64}.  As far as I know, neither he nor Einstein explain why, exactly, such violation is supposed to be objectionable.  At the very least, given Newton's open-eyed advocacy of absolute space, it seems peculiar to describe it as ``contrary to the mode of scientific thinking.''}  It is clear that, while GR fails to fulfil the Machian goal of providing a reductive account of the local inertial frames, it does not suffer from this newly identified (alleged) defect of pre-relativistic theories.  The metric structure of GR conditions the evolution of the material content of spacetime, but it is also, in turn, affected by that content.

This potted review of Einstein's early pronouncements is intended to show that he was one of the original advocates of the view outlined in Section~\ref{s:special}, namely, that GR differs from its predecessors, not through lacking the kind of spacetime structures that such theories have, but by no longer treating that structure as a non-dynamical background.  It also shows that, despite being responsible for the idea that the general covariance of GR has physical significance as the expression of the theory's generalisation of the relativity principle, Einstein himself quickly retreated from this idea.  He continued (mistakenly) to espouse the idea that GR generalised the principle of relativity, via the equivalence principle, but GR's general covariance was no longer taken to be a sufficient condition of its doing so.  Instead the implication in the opposite direction was stressed.  General covariance was taken to be a \emph{necessary} condition of implementing a general relativity principle: there can be no special coordinate systems adapted to preferred states of motion in a theory in which there are no preferred states of motion!

In the immediate wake of Kretschmann's criticism, one of Einstein's most revealing statements concerning the status of general covariance comes in his response to a paper by Ernst Reichenb\"{a}cher.  There Einstein contrasts a theory that includes an acceleration standard with one that does not:
\begin{quotation}
if acceleration has absolute meaning, then the nonaccelerated coordinate systems are preferred by nature, i.e., the laws then must---when referred to them---be different (and simpler) than the ones referred to accelerated coordinate systems. Then it makes no sense to complicate the formulation of the laws by pressing them into a generally covariant form.

Vice versa, if the laws of nature are such that they do not attain a preferred form through the choice of coordinate systems of a special state of motion, then one cannot relinquish the condition of general covariance as a means of research. \citep[205]{einstein20}
\end{quotation}
From a modern perspective, several things are notable about this passage.  First, GR qualifies as a theory whose laws do not attain a ``preferred form through the choice of coordinate systems of a special state of motion,'' not because (as Einstein believed) acceleration does not have an absolute meaning in the theory, but because the structure that defines absolute acceleration is no longer homogeneous; in general, it is not possible to define, over a neighbourhood of a point in spacetime, a coordinate system whose lines of constant spatial coordinate are both non-accelerating absolutely and not accelerating with respect to each other.  GR lacks a non-generally-covariant formulation,\footnote{\label{n:fock}Even this can be disputed.  Fock, for example, argued that harmonic coordinates, defined via the condition $(g^{\mu \nu}\sqrt{-g})_{, \mu} = 0$, have a preferred status in GR, analogous to that of Lorentz charts in special relativity.} but not for the reason Einstein suggests.

Second, while the equations expressing a theory's laws might be \emph{simpler} in a coordinate system adapted to the theory's standard of acceleration, it does not follow that these equations, and the equations that hold with respect to accelerated coordinate systems, express different laws.  In fact, it is much more natural to see the formally different equations as but different coordinate-dependent expressions of the same relations holding between coordinate-independent entities.  As Anderson says of entities that occur explicitly in a generally covariant formulation of some laws but which were not apparent in the non-(generally)-covariant equations: ``these elements were there in the first place, although their existence was masked by the fact that they had been assigned particular values. That is, the $g^{\mu \nu}$ [of a generally covariant formulation of a special relativity] are present in [the Lorentz-covariant form of] special relativity with the fixed preassigned values of the Minkowski metric'' \citep[192]{anderson64}.\footnote{\label{ATF}The same view of the meaning of the preferred coordinates of the non-covariant form of Newtonian gravitation theory is clearly articulated by \citet[418]{trautman66}.  It was thoroughly assimilated in the philosophical literature; see, e.g., \citet[54--55]{friedman83}.  The perspective is explored further in Sections~\ref{s:gcvsdi} and \ref{s:coords}, where I argue that its relevance for discussions of alleged differences between the observables of GR and pre-relativistic theories has not been fully appreciated.}

Finally, while calculation might not be aided by complicating the formulation of the laws by expressing them generally covariantly, conceptual clarity can be.  Real structures that are only implicit in the non-covariant formalism are laid bare in the generally-covariant formalism, and their status can then be subjected to scrutiny.

In fact, Einstein himself says something quite consonant with these observations earlier in the same paper:
\begin{quote}
the coordinate system is only a \emph{means of description} and in itself has nothing to do with the \emph{objects to be described}. Only a law of nature in a generally covariant form can do complete justice in this situation, because in any other way of describing, statements about the means of description are jumbled with statements about the object to be described. \citep[203]{einstein20}
\end{quote}
Einstein's idea seems to be that coordinates should not have a function beyond the mere labelling of physical entities, the qualitative character of which is to be fully described by other means.  But this is a basis, not for an argument in favour of laws that can only be expressed generally covariantly (seemingly Einstein's intention), but for an argument for the generally-covariant formulation of laws in general, whatever they be.  Ironically, it is an argument that is most relevant to pre-relativistic theories, not GR, because only in this context can one choose to encode physically meaningful quantities (spacetime intervals) via special choices of coordinate system, and thereby `jumble up' the mode of description with that described.

\section{Dissent from Quantum Gravity}\label{s:dissent}

Let me sum up the picture presented so far.  General covariance \emph{per se} has no physical content:  the essence of Kretschmann's objection to Einstein is that any sensible theory can be formulated in a generally covariant manner.  It follows that GR does not differ from SR in virtue of having a generally covariant formulation.  However, GR does differ from SR in \emph{lacking} a \emph{non}-covariant formulation. 
\label{n:bergmann} Some authors have made this fact the basis for claiming that GR, but not SR, satisfies a ``principle of general covariance''.  For example, Bergmann writes: ``The hypothesis that the geometry of physical space is represented best by a formalism which is covariant with respect to general coordinate transformations, and that a restriction to a less general group of transformations would not simplify that formalism, is called \emph{the principle of general covariance}'' \citep[159]{bergmann42}.

In SR the existence of a non-covariant formulation is connected with the failure of a general principle of relativity.  The privileged coordinate systems of SR, in which the equations expressing the laws simplify, encode (\emph{inter alia}) a standard of non-accelerated motion.  There can be no preferred coordinate systems (of such a type) in a theory that implements a general principle of relativity.  This might suggest that GR's lack of a non-covariant formulation is connected to the generalisation of a relativity principle, but (\emph{pace} Einstein) it stems from no such thing.  Rather, the lack of preferred coordinates is due to the fact that the spacetime structures of a generic solution, including those structures common to SR and GR that define absolute acceleration (in essentially the same way in both theories), lack symmetries and so cannot be encoded in special coordinates.

Finally, this lack of symmetry is entailed by, \emph{but does not entail}, the fundamental distinguishing feature of GR, namely, that the structure encoded by the metric of GR is, unlike that of SR, dynamical.  A fully dynamical field, free to vary from solution to solution, will generically lack symmetries.  So a background independent theory, in which all fields are dynamical, will lack a non-covariant formulation (of the relevant kind).  The converse, however, is not true.  In principle we can define a theory involving a background metric with no isometries, and such a theory will only have a generally covariant formulation.\footnote{Smolin demurs: ``if one believes that the geometry of space is going to have an absolute character, fixed in advance, by some a priori principles, you are going to be led to posit a homogeneous geometry. For what, other than particular states of matter, would be responsible for inhomogeneities in the geometry of space?'' \citep[201]{smolin06}. But why does a background geometry need to be fixed by ``a priori principles''?  Its being what it is could simply be brute fact, inhomogeneities notwithstanding.}
%

Something like this collection of commitments, though not uncontroversial, represents a mainstream view, at least amongst more recent textbooks in the tradition of \citet{synge60} and \citet{mtw73}.  Unfortunately, there is a fly in the ointment, for it apparently conflicts with a dominant view amongst many in the quantum gravity community, in particular, the founding fathers of loop quantum gravity.  Workers in this field often endorse the idea that GR's background independence, understood as the absence of `fixed', non-dynamical spacetime structure, is its defining feature.  But they go on to link this property to the theory's general covariance, or, to use the more favoured label, its \emph{diffeomorphism invariance}.  For example, Lee Smolin claims that ``both philosophically and mathematically, it is diffeomorphism invariance that distinguishes general relativity from other field theories'' \cite[234]{smolin03}. 

 And Carlo Rovelli, who has perhaps written the most on the link between background independence and diffeomorphism invariance, says of the background independence of classical GR that ``technically,
it is realized by the gauge invariance of the action under (active)
diffeomorphisms'' \cite[10]{rovelli04}, and (perhaps in less careful moments) he treats the two as synonymous \citep[279]{rovelligaul00}.

On the face of it, these claims conflict with the Kretschmann view.  They appear to assert that a formal property of GR, its ``(active) diffeomorphism invariance'', has physical content in virtue of realising, or expressing, a physical property of the theory, namely, its background independence.  Since specially relativistic theories are not background independent (as we have been understanding this term), it should follow that they cannot be formulated in a diffeomorphism invariant manner.  At the very least, if one follows Kretschmann in supposing that any theory can be formulated in a generally covariant manner, then (active) diffeomorphism invariance, as understood by Rovelli \emph{et al.}, cannot be the same as general covariance as understood in the Kretschmann tradition.  And, indeed, the same authors routinely draw distinctions of this kind.

Much of the rest of this paper is concerned to see how far one can push back against the Rovelli--Smolin line, in the spirit of Kretschmann and Friedman.  What the exercise reveals is that the connection between diffeomorphism invariance and background independence is messier, and less illuminating, than recent discussions originating in the quantum gravity literature might suggest.  It also sheds light on a different but closely related topic.  In the same discussions, the diffeomorphism invariance and/or background independence of GR is frequently taken to have profound implications for the nature of the theory's observables.  It is important that a merely technical sense of ``observable'' is not all that is at issue.  The claim often appears to be that GR and pre-relativistic theories differ in terms of the kind of thing that is observable in a non-technical sense.  In other words, it is alleged that the theories differ over the fundamental nature of the physical magnitudes that they postulate.\footnote{Amongst philosophers, \citet{earman06b} and \citet{rickles08} are proponents of variants of this view.}  This, I believe, is a mistake, as I hope some of the distinctions to be reviewed below help to show.

The first task is to clarify what might be meant by ``diffeomorphism invariance'' as distinct from ``general covariance''.  I then revisit the notion of a background field, as characterised informally above, for finer-grained distinctions should be drawn here too.

\section{General Covariance vs Diffeomorphism Invariance}\label{s:gcvsdi}

Several authors have drawn what they presumably take to be the crucial, bipartite distinction between types of general covariance and diffeomorphism invariance.  Norton, for example, distinguishes ``active'' and ``passive'' general covariance \citep[1226, 1230]{nortonCC}.  Rovelli distinguishes ``active diff invariance'' from ``passive diff invariance'' \citep[122]{rovelli01}.  Earman distinguishes merely ``formal'' from ``substantive'' general covariance \citep{earman06,earman06b}.  Ohanian and Ruffini distinguish ``general covariance'' from ``general invariance'' \citep[276--9]{ohanianruffini13}.  Finally, Giulini distinguishes ``covariance under diffeomorphisms'' from ``invariance under diffeomorphisms'' \citep[108]{giulini07}.  As this cornucopia of terminology indicates, several different distinctions are in play, and linked to further ancillary notions (for example, that between ``active'' and ``passive'' transformations) in myriad ways.  In the face of this morass, my strategy will be to articulate as clearly as I can what I take to be the most useful distinction, before relating it to several of the ideas just listed.

In differentiating distinct notions of general covariance and diffeomorphism invariance, it will be useful to consider various concrete formulations of theories that exemplify the properties in question.  Further, when contrasting specially and generally relativistic theories, it is good policy to eliminate unnecessary and potentially misleading differences by choosing theories that are as similar as possible.  My running example, for both the specially and generally relativistic cases, will be theories of a relativistic massless real scalar field, $\Phi$. 

In the context of SR, such a field obeys the Klein--Gordon equation, but there are at least three ``versions'' of this equation to consider:
\begin{gather}
\frac{\partial^2\Phi}{\partial x^2} +
\frac{\partial^2\Phi}{\partial y^2} + \frac{\partial^2\Phi}{\partial
z^2} - \frac{\partial^2\Phi}{\partial t^2} = 0, \label{kgsr} \\
\eta^{\mu \nu}\Phi_{;\nu \mu} = 0, \label{kgsrgc}\\
\eta^{ab}\nabla_a \nabla_b \Phi = 0. \label{kgsrci}
\end{gather}
These equations are most plausibly understood as (elements of) different formulations of one and the same theory, not as characterising different theories. This requires that the equations are understood as but different ways of picking out the very same set of models (and thereby the very same set of physical possibilities).  On the picture that allows this, one also gains a better understanding of the content of each equation.

What is that picture?  Start with equation~\eqref{kgsrci}.  The roman indices occurring in the equation are ``abstract indices'', indicating the type of geometric object involved.  This equation, therefore, is not to be interpreted (as the other two are) as relating the coordinate components of various objects.  Rather, it is a direct description of (the relations holding between) certain geometric object fields defined on a differentiable manifold.  Its models are triples of the form $\langle M, \eta_{ab}, \Phi \rangle$: differential manifolds equipped with a (flat) Lorentzian metric field $\eta_{ab}$ and a single scalar field $\Phi$. (I am taking the torsion-free, metric-compatible derivative operator, $\nabla$, to be defined in terms of the metric field; it is not another primitive object, over and above $\eta_{ab}$ and $\Phi$.)

Equations~\eqref{kgsr} and \eqref{kgsrgc} are to be understood as ways of characterising the very same models, but now given under certain types of coordinate description.  In particular, in the case of equation~\eqref{kgsr}, one is choosing coordinates that are specially adapted to symmetries of one of the fields of the model, namely, the flat Minkowski metric.  Such coordinates are singled out via the ``coordinate condition'' $\eta_{\mu \nu} = \text{diag}(-1,1,1,1)$.   In the case of equation~\eqref{kgsrgc}, one is allowing any coordinate system adapted to the differential structure of the manifold, $M$.

We are now in a position to draw the crucial distinction between general covariance (as it has been implicitly understood in the previous sections) and diffeomorphism invariance for, on one natural way of further filling in the details, although it is generally covariant, \emph{the theory just given fails to be diffeomorphism invariant}.

First, general covariance.  We define this as follows:
\begin{GC}
A formulation of a theory is \emph{generally covariant} iff the equations expressing its laws are written in a form that holds with respect to all members of a set of coordinate systems that are related by smooth but otherwise arbitrary transformations.
\end{GC}
It is clear that such a formulation is possible for our theory.  It is what is achieved in the passage from the traditional form of the equation, \eqref{kgsr}, to equation~\eqref{kgsrgc}.  General covariance in this sense is sometimes taken to be equivalent to the claim that the laws have a \emph{coordinate-free formulation} (\citealp[54]{friedman83}; \citealp[108]{giulini07}). This takes us to equation~\eqref{kgsrci}: if the laws relate geometric objects of types that are intrinsically characterisable, without recourse to how their components transformations under changes of coordinates,  then one should be able, with the introduction of the right notation, to describe the relationships between them directly, rather than in terms of relationships that hold between the objects' coordinate components.

In order to address the question of the theory's diffeomorphism invariance, one needs to be more explicit than we have so far been about how one should understand equation~\eqref{kgsrci}.  In particular, what, exactly, is the referent of the `$\eta_{ab}$' that occurs in this equation?  Here is one very natural way to set things up.  It is a picture that lies behind the claim of several authors that, while specially relativistic theories can be made generally covariant in the sense just described, they are nevertheless \emph{not} diffeomorphism invariant.

Take the \emph{kinematically possible models} (KPMs) of the theory to be suitably smooth functions from some given manifold equipped with a Minkowski metric, $\langle M, \eta_{ab} \rangle$ into $\mathbb{R}$.  That is, they are objects of the form $\langle M, \eta_{ab}, \Phi \rangle$, where $\eta_{ab}$ is held \emph{fixed}---it is \emph{identically} the same in every model.\footnote{This means that the concept of a fixed field is not equivalent to the concept of an absolute object in the Anderson--Friedman sense. In using ``fixed'' in this quasi-technical sense, I follow \citet[see, e.g.,][197, fn~137]{belot07}. The distinction is explored more fully in Section~\ref{s:AO}.}  The \emph{dynamically possible models} (DPMs) are then the proper subset of these objects picked out by the requirement that $\Phi$ satisfies the Klein--Gordon equation relative to the $\eta_{ab}$ common to all the KPMs.  So understood, equation~\eqref{kgsrci} is not an equation for $\eta_{ab}$ and $\Phi$ together.  Rather, it is an equation for $\Phi$ alone, \emph{given} $\eta_{ab}$ \citep[\emph{cf.}][107]{giulini07}.  For ease of future reference, call this version of the specially relativistic theory of the scalar field \textbf{SR1}.

Our initial definition of diffeomorphism invariance runs as follows:
\begin{DI1} A theory $T$ is \emph{diffeomorphism invariant} iff, if $\langle M, O_1, O_2, \ldots \rangle$ is a solution of $T$, then so is $\langle M, d^*O_1, d^*O_2, \ldots \rangle$ for all $d \in \text{Diff}(M)$.\footnote{In this statement of the condition, $O_i$ and $d^*O_i$ are distinct mathematical objects; one is not contrasting different coordinate representations of the very same objects.}
\end{DI1}
So defined, diffeomorphism invariance corresponds to what has sometimes simply been identified as general covariance in the post-Hole Argument philosophical literature.\footnote{%
See, e.g., \citet[47]{earman89}.  As mentioned, Norton distinguishes active and passive general covariance.  His statement of the former \citep[1226]{nortonCC} is almost identical to the statement of diffeomorphism invariance just given, save that he considers diffeomorphisms between distinct manifolds.  (His statement of passive general covariance \cite[1230]{nortonCC} differs, however, from the characterisation of general covariance given above, in focusing on the closure properties of the set of coordinate representations of a theory's models, rather than on the nature of the equations that pick out such models.)%
}  Friedman is explicit in taking general covariance as defined above \citep[\emph{cf.}][51]{friedman83} to be equivalent to diffeomorphism invariance as just defined \citep[\emph{cf.}][58]{friedman83}.  In arguing for this equivalence \citep[52--4]{friedman83}, he appears to overlook the crucial possibility, exploited here, that a coordinate-free equation relating two geometric objects $A$ and $B$, can nonetheless be interpreted as an equation for $B$ alone, given a fixed $A$.  (We shall see in Section~\ref{s:gauge} that \citet{earman06} seems to be guilty of a similar oversight.)

Returning to \textbf{SR1}, it is clear that, with the KPMs and DPMs defined as suggested, the theory does \emph{not} satisfy the definition of diffeomorphism invariance just given.  If $\langle M, \eta_{ab}, \Phi \rangle$ is a model of the theory, $\langle M, d^*\eta_{ab}, d^*\Phi \rangle$ will be a model only if $d^*\eta_{ab} = \eta_{ab}$, for only in that case will $\langle M, d^*\eta_{ab}, d^*\Phi \rangle$ correspond to a KPM, let alone a DPM!

Contrast \textbf{SR1} to the generally relativistic theory of the scalar field.  To make the analogy as close as possible, consider the sector of the theory defined on the same manifold $M$ mentioned in \textbf{SR1}. Call this theory \textbf{GR1}.  Superficially, the KPMs and the DPMs of \textbf{GR1} are the same type of objects as those of \textbf{SR1}: triples of the form $\langle M, g_{ab}, \Phi \rangle$, where $g_{ab}$, like $\eta_{ab}$, is a Lorentzian metric field.  But now one does not have the option of taking $g_{ab}$ to be fixed.\footnote{Strictly speaking, one could interpret equations~\eqref{kggr} and \eqref{kgEFE}, given below, as describing a theory of a single field $\Phi$ propagating on a fixed $g_{ab}$.  The resulting space of DPMs would consist of a single point in this cut-down space of KPMs!  What, exactly, would be wrong with such a setup?  We take ourselves to have evidence for the (approximate) truth of our theory (GR) even though we have not pinned down a specific model.  But on this variant of the theory, pinning down the theory requires pinning down a unique model.}  Rather the KPMs of the theory are \emph{all possible} triples of the form $\langle M, g_{ab}, \Phi \rangle$, subject only to $g_{ab}$ and $\Phi$ satisfying suitable differentiability (and perhaps boundary) conditions.  The DPMs are picked out as a proper subset of the KPMs by two equations:
\begin{gather}
g^{ab}\nabla_a \nabla_b \Phi = 0, \label{kggr} \\
G_{ab} = 8\pi T_{ab}. \label{kgEFE}
\end{gather}

Equation~\eqref{kgEFE} is the Einstein field equation, relating the Einstein tensor $G_{ab}$, encoding certain curvature properties of $g_{ab}$, to the energy momentum tensor $T_{ab}$.\footnote{For our massless real scalar field, $T_{ab} = (\nabla_{\!a}\Phi) (\nabla_{\!b}\Phi) - \frac{1}{2}g_{ab}g^{mn}(\nabla_{\! m} \Phi) (\nabla_{\! n}\Phi) $
.}  Equation~\eqref{kggr} might look superficially like equation~\eqref{kgsrci}, but now it is no longer an equation for $\Phi$ given $g_{ab}$.  Rather \eqref{kggr} and \eqref{kgEFE} together form a coupled system of equations---the ``Einstein--Klein--Gordon equations''---for $g_{ab}$ and $\Phi$ together.  This generally relativistic theory is, of course, diffeomorphism invariant: if $\langle M, g_{ab}, \Phi \rangle$ satisfies equations~\eqref{kggr} and \eqref{kgEFE}, so does $\langle M, d^*g_{ab}, d^*\Phi \rangle$ for any diffeomorphism $d$.

The rather dramatic way in which \textbf{SR1} fails to meet our definition of diffeomorphism invariance---that for a generic diffeomorphism $d$, $\langle M, d^*\eta_{ab}, d^*\Phi \rangle$ is not even a KPM when $\langle M, \eta_{ab}, \Phi \rangle$ is a DPM---suggests a modification of our definition.  Rather than considering the effect of a diffeomorphism on all of the fields of a theory's models, we can exploit the distinction, built into the very construction of the theory, between fixed fields and dynamical fields.  Letting $F$ stand for the solution-independent fixed fields common to all KPMs, and letting $D$ stand for the dynamical fields, we can consider the effect of acting only on the latter.  This leads to the following amended definition:
\begin{DI2}
A theory $T$ is \emph{diffeomorphism invariant} iff, if $\langle M, F, D \rangle$ is a solution of $T$, then so is $\langle M, F, d^*D \rangle$ for all $d \in \text{Diff}(M)$.
\end{DI2}
\noindent More generally, one can say that a theory $T$ is $G$-invariant, for some subgroup $G \subseteq \text{Diff}(M)$ iff, if $\langle M, F, D \rangle$ is a solution of $T$, then so is $\langle M, F, g^*D \ldots \rangle$ for all $g \in G$.

Since \textbf{GR1} involves no fixed fields, acting only on the dynamical fields \emph{just is} to act on all the fields.  Our amendment to the definition of diffeomorphism invariance therefore makes no material difference in this case.  For this reason, focus on theories like \textbf{GR1} tends to obscure the difference between our two definitions.  Turning to the case of \textbf{SR1}, this theory still fails to be diffeomorphism invariant under the new definition: for an arbitrary diffeomorphism $d$, if $\langle M, \eta_{ab}, \Phi \rangle$ is a solution of \textbf{SR1}, then $\langle M, \eta_{ab}, d^*\Phi\rangle$, in general, will not be.  However, assuming no boundary conditions are being imposed, $\langle M, \eta_{ab}, d^*\Phi\rangle$ will nonetheless be a KPM of the theory.  This becomes significant when considering the definition of the invariance of the theory under proper subgroups of $\text{Diff}(M)$.

Suppose $T$ has models of the form $\langle M, F, D \rangle$ and that $d$ is a symmetry of the fixed, background structure, i.e., $d^*F = F$.  In this case, $\langle M, d^*F, d^*D \rangle = \langle M, F, d^*D \rangle$ and so, for this subgroup of $\text{Diff}(M)$, an invariance principle that asks us to consider transformations of all fields, background and dynamical, will give the same verdict as those that consider transformations only of the dynamical fields.  Further, it follows from the general covariance of the theory, i.e., from the fact that its defining equation can be give a coordinate-free expression, that when $d$ is a symmetry of $F$, $\langle M, d^*F, d^*D \rangle = \langle M, F, d^*D \rangle$ will be a DPM whenever $\langle M, F, D \rangle$ is.\footnote{Note that this claim is not identical to Earman's claim that it follows from general covariance that a diffeomorphism that is symmetry of a theory's spacetime structure will also be what he calls a ``dynamical symmetry'' \citep[46--7]{earman89}.  The reason is that Earman's ``general covariance'' corresponds to the (unmodified) definition of diffeomorphism invariance given above.}  We can therefore define $G$-invariance either by analogy with the first definition of diffeomorphism invariance or (as advocated) by analogy with the final version, and we will get the verdict that if $G$ is a subgroup of the automorphism group of $F$, then the theory is $G$-invariant.

The definitions give different verdicts, however, when we consider the opposite implication: if $T$ is a $G$-invariant theory, does it follow that $G$ is a subgroup of the automorphism group of its fixed fields $F$?  If $G$-invariance requires that if $\langle M, F, D \rangle$ is a DPM then so is $\langle M, g^*F, g^*D \ldots \rangle$, for all $g \in G$, then no diffeomorphism that is not also an automorphism of $F$ could be a member of $G$.  Such a diffeomorphism does not map KPMs to KPMs.  However, if $G$-invariance only requires that if $\langle M, F, D \rangle$ is a DPM then so is $\langle M, F, g^*D \ldots \rangle$, then the automorphisms of $F$ can be a \emph{proper} subgroup of $G$.  In fact, this is exactly the situation in the case of \textbf{SR1}.  Let $d$ correspond to a conformal transformation of $\eta_{ab}$.  Since we are considering the \emph{massless} Klein--Gordon field, if $\langle M, \eta_{ab}, \Phi\rangle$ is a DPM, then so is $\langle M, \eta_{ab}, d^*\Phi \rangle$, even though $d^*\eta_{ab} \neq \eta_{ab}$.  We can only capture this fact in terms of the statement that the theory is invariant under the relevant group if we define such invariance in the modified manner.\footnote{Similar, historically-inspired examples are Galilean-invariant classical mechanics set in full Newtonian spacetime and, more interestingly, Newtonian gravitational theory set in Galilean spacetime \citep[see, e.g.,][]{knoxGNG}.  What these examples should remind one is that such theories are epistemologically problematic. The background structure that they postulate introduces allegedly meaningful properties (e.g., absolute velocities) that are undetectable in principle.  This motivates the search for formulations with weaker background structure (\citealp[see, e.g.][\S3 and \S6]{pooleysubrel}).}

Let's take a step back and recall the wider project.  We are interested in assessing the claim that  diffeomorphism invariance is intimately linked to background independence.  I contend that the distinction drawn in this section between general covariance and diffeomorphism invariance, and exemplified by \textbf{SR1}'s satisfaction of the first but not the second, is the right one for this purpose, for it makes good sense of several remarks by the claim's defenders.

For example, \citet[\S6]{smolin03} offers an extended discussion of diffeomorphism invariance and its connection to background independence.  His focus is on the interpretational consequences of diffeomorphism invariance, rather than on providing a positive characterisation of the property as such, so no direct comparison with the definition proposed here can be made. (He is also particularly concerned to stress the \emph{gauge} status of diffeomorphisms in the context of a diffeomorphism-invariant formulation of a theory, a topic I return to in Section~\ref{s:gauge}.)  However, his contrasting diffeomorphism invariance with general coordinate invariance is fully consonant with the distinction of this section:
\begin{quote}
it can be asserted---indeed it is true---that with the introduction of explicit background fields any field theory can be written in a way that is generally coordinate invariant. This is not true of diffeomorphisms [sic] invariance, which relies on the fact that in general relativity there are no non-dynamical background fields. \cite[233]{smolin03}
\end{quote}
It is natural to read the second half of this passage as committing Smolin to the claim that \textbf{SR1} cannot be made diffeomorphism invariant because the theory involves a non-dynamical background, $\eta_{ab}$.

Consider, now, a revealing passage from Rovelli.  Having summarised what he takes to be the philosophical implications of GR's lack of non-dynamical background structures, he states that these implications are ``coded in the active diffeomorphism invariance (diff invariance) of GR'' \cite[108]{rovelli01}.  He goes on to elaborate in a footnote:
\begin{quote}
Active diff invariance should not be confused with passive diff invariance, or invariance under change of co-ordinates\ldots
A field theory is formulated in [a] manner invariant under passive diffs (or change of co-ordinates), if we can change the co-ordinates of the manifold, re-express all the geometric quantities (dynamical \emph{and non-dynamical}) in the new coordinates, and the form of the equations of motion does not change. A theory is invariant under active diffs, when a smooth displacement of the dynamical fields (\emph{the dynamical fields alone}) over the manifold, sends solutions of the equations of motion into solutions of the equations of motion. \citep[122]{rovelli01}
\end{quote}
I take it that \textbf{SR1} is precisely a theory formulated in a manner invariant under passive diffs, but not active diffs, whereas \textbf{GR1} is a theory invariant under active diffs.  In other words, Rovelli's ``passive diffeomorphism invariance'' is what I called above general covariance.  Identifying Rovelli's ``non-dynamical'' fields with fixed fields, his ``active diffeomorphism invariance'' corresponds to our (amended) definition of diffeomorphism invariance.

Finally, \citet{giulini07} offers equivalent definitions, although he adopts a rather different approach to characterising general covariance.  He schematically represents a theory's equations of motion as:
\begin{equation}
\mathcal{F}[\gamma, \Phi, \Sigma] = 0 \label{DGeqn}
\end{equation}
Here $\gamma$ goes proxy for structures given by maps \emph{into} the manifold $M$  (representing particle worldlines, strings etc.)\ and $\Phi$ goes proxy for the dynamical fields: maps from spacetime into some value space (or, more generally, structures given by sections in some bundle over $M$).  Finally, $\Sigma$ stands for the fixed (``background'') structures.\footnote{In both our examples theories, $\gamma$ is empty and the scalar field $\Phi$ belongs to Giulini's category $\Phi$.  In the case of \textbf{SR1}, $\eta_{ab}$ belongs to $\Sigma$; in \textbf{GR1}, $g_{ab}$ belongs to (Giulini's) $\Phi$, and $\Sigma$ is empty.}

He then distinguishes what he calls the notion of \emph{covariance} from \emph{invariance} as follows \citep[see][108]{giulini07}.  Equation~\eqref{DGeqn} is said to be \emph{covariant} under diffeomorphisms iff:
\begin{equation}
\mathcal{F}[\gamma, \Phi, \Sigma] = 0 \quad \text{iff} \quad\mathcal{F}[d\cdot \gamma, d\cdot \Phi, d\cdot \Sigma] = 0 \quad \forall d \in \text{Diff}(M). \label{DGcov}
\end{equation}
It is \emph{invariant} under diffeomorphisms iff:
\begin{equation}
\mathcal{F}[\gamma, \Phi, \Sigma] = 0 \quad \text{iff} \quad\mathcal{F}[d\cdot \gamma, d\cdot \Phi, \Sigma] = 0 \quad \forall d \in \text{Diff}(M). \label{DGinv}
\end{equation}
The only difference between these conditions is that in the former but not in the latter case one allows the diffeomorphism to act on the fixed fields.  In absence of fixed fields, therefore, the distinction between the conditions collapses:  covariance implies invariance.

The distinction between the $\gamma$ and $\Phi$, on the one hand, and the $\Sigma$ on the other is crucial in understanding these conditions.  Consider, first, condition~\eqref{DGinv}.  The statement that $\mathcal{F}[\gamma, \Phi, \Sigma] = 0$  iff $\mathcal{F}[d\cdot \gamma, d\cdot \Phi, \Sigma] = 0$ simply means that $\langle \gamma, \Phi \rangle$ and $\langle d\cdot \gamma, d\cdot \Phi \rangle$ stand or fall together as solutions of \eqref{DGeqn}.  The condition is therefore this section's (modified) statement of diffeomorphism invariance.

Now consider condition~\eqref{DGcov}.  The fact that $\mathcal{F}[\gamma, \Phi, \Sigma] = 0$ is only an equation for $\gamma$ and $\Phi$ (but not $\Sigma$) means that $\mathcal{F}[\gamma, \Phi, \Sigma] = 0$ and $\mathcal{F}[d\cdot \gamma, d\cdot \Phi, d\cdot \Sigma] = 0$ are \emph{distinct} equations.  The condition states that if $\langle \gamma, \Phi \rangle$ is a solution to \eqref{DGeqn}, then $\langle d\cdot \gamma, d\cdot \Phi \rangle$ must be a solution of a structurally similar equation involving the \emph{different} field(s) $d \cdot \Sigma$.  The condition \eqref{DGcov}, therefore, says nothing about whether $d$ maps a solution of \eqref{DGeqn} to another solution \emph{of the same equation}.  Given that $\Sigma$ represents fixed fields, \eqref{DGcov} does not collapse into our original, unmodified statement of diffeomorphism invariance.  All that it requires is that \eqref{DGeqn} be well defined in the differential-geometric sense.  It is therefore equivalent to the requirement that the equation have a generally covariant expression in the sense given earlier.


\section{Diffeomorphism-Invariant Special Relativity}

The previous section described a generally-covariant but non-diffeomorphism-invariant formulation of an intuitively background-dependent theory, \textbf{SR1}.  This was contrasted with a generally-covariant and diffeomorphism-invariant formulation of an intuitively background-independent theory, \textbf{GR1}.\footnote{From here on, when I refer simply to ``diffeomorphism invariance'' I am referring to the property captured by the second (final) definition given in the previous section.  The merits, or otherwise, of the first definition will not be discussed further.}  What should one make of \textbf{SR1}'s failure to be diffeomorphism invariant?  Does it support Smolin's contention that diffeomorphism invariance ``relies on'' the absence of background fields?  In this section and the next, I suggest that it does not.  At the very least, whether it does depends on what counts as a ``background field.''

We need to consider yet another formulation of a theory, which I'll call \textbf{SR2}.  This theory's space of KPMs is the very same set of objects that formed the space of KPMs of the generally-relativistic \textbf{GR1}.  But, rather than being picked out via equations~\eqref{kggr} and \eqref{kgEFE}, the subspace of DPMs is defined via:
\begin{gather}
g^{ab}\nabla_a \nabla_b \Phi = 0, \tag{\ref{kggr}}\\
R^{a}_{\phantom{a}bcd} = 0, \label{riem0}
\end{gather}
where $R^{a}_{\phantom{a}bcd}$ is the Riemann curvature tensor of $g_{ab}$.\footnote{As with those of \textbf{GR1}, the theory's KPMs are restricted to fields defined on a given manifold $M$.  In the previous section, this restriction served to allow as direct as possible a comparison between \textbf{GR1} and \textbf{SR1}.  When comparison with \textbf{SR1} is not at issue, the restriction is arbitrary.  One can (and should) generalise the formulations of \textbf{SR2} and \textbf{GR1} further, not least to allow for different global topologies.}  Several comments are in order before we assess the interpretational dilemmas that \textbf{SR2} presents.

First, the contrast between \textbf{SR1} and \textbf{SR2} highlights something of a contrast between the philosophy literature, including the post-Hole Argument literature, and discussions of background independence arising from attempts to quantise GR.  Crudely put, philosophers have tended to have a formulation of a theory like \textbf{SR2} in mind when they have considered `generally covariant' formulations of special relativity \citep[see, e.g.,][518]{earmannorton87}, whereas physicists have tended to have something like \textbf{SR1} in mind.  This is not unrelated to the fact, noted in the previous section, that Friedman, Earman, and even Norton (used to) identify (active) general covariance with diffeomorphism invariance (as initially characterised in the previous section).

This is not to say that the physics literature has not discussed theories like \textbf{SR2}---we shall shortly see that it has---but it is possible to mistake a discussion of an \textbf{SR1}-type theory for that of a \textbf{SR2}-type theory. 
 One does not arrive at \textbf{SR2} simply by stipulating that equation~\eqref{riem0} is to be satisfied.  One must also indicate how $g_{ab}$, as it occurs in \eqref{kggr} and \eqref{riem0}, is to be interpreted.  After all, the field $\eta_{ab}$ of \textbf{SR1} satisfies a formally identical equation to \eqref{riem0}.  It is just that, in this context, the equation does not function to pick out a class of DPMs from a wider class of KPMs.  Instead it characterises a fixed field common to all the KPMs.  In \textbf{SR2}, it is important that \eqref{kggr} and \eqref{riem0}, just like \eqref{kggr} and \eqref{kgEFE} in \textbf{GR1}, are understood as coupled equations for both $\Phi$ \emph{and} $g_{ab}$.

Finally, of course, we should note the crucial fact that \textbf{SR2}, like \textbf{GR1} and unlike \textbf{SR1}, is diffeomorphism invariant.

\section{Connecting Diffeomorphism Invariance and Background Independence}\label{s:connect}

What does the the diffeomorphism invariance of \textbf{SR2} tell us about the alleged link between diffeomorphism invariance and background independence?  A proper answer to this question will require disentangling various meanings of ``background'', but here is the obvious moral: \textbf{SR2} is a diffeomorphism-invariant but intuitively background-dependent theory.  Diffeomorphism invariance therefore cannot be equated with---or be seen as a formal expression of, or sufficient condition for---background independence. Diffeomorphism invariance is not, \emph{per se}, what differentiates GR from pre-relativistic theories.

Here is one way that this conclusion might be resisted.  Consider the following questions.  (Q1) Is \textbf{SR2} a background-independent theory?  (Q2) Are \textbf{SR1} and \textbf{SR2} merely different ways of formulating the same theory?  Suppose that one answers (Q1) in the affirmative, on the grounds that $g_{ab}$ in a model of the theory is a \emph{solution to an equation}.  It therefore counts as a `dynamical field'; it is not `fixed a priori'.  This, in effect, is to treat `background field' as synonymous with `solution-independent fixed field' in the sense highlighted in Section~\ref{s:gcvsdi}.  One then goes on to answer question (Q2) in the negative.  Precursors of GR were not background independent, period, and so only \textbf{SR1} is faithful to the pre-GR understanding of the spacetime structure of special relativity.

I take it that this package is a highly implausible cocktail of views.  First, one should ask: on what basis can one assert that \textbf{SR1} and \textbf{SR2} constitute genuinely distinct theories, rather than merely different formulations of the same theory?  On the face of it, since their models involve the same types of geometric object, and since all objects in any solution of one theory are diffeomorphic to the corresponding objects in some solution of the other, the two formulations appear to be, not merely empirically equivalent, but equivalent in a thoroughgoing sense.  The DPMs of one theory are isomorphic to the DPMs of the other; it is just that, for each solution of one of the theories, the other theory has an infinite set of diffeomorphic copies.

Second, the classification of \textbf{SR2} as relevantly similar to \textbf{GR1}, and so background independent, focuses on a minor similarity between the theories at the expense of a more significant contrast.  True, the $g_{ab}$s of both theories  are treated as `solutions of equations' and \emph{in this sense} they are not fixed, but this fact seems much less interesting than their obvious differences.  Recall the intuitive characterisation of the differences between the spacetime structures of GR and pre-relativistic theories given in Section~\ref{s:special}: in GR, the curvature of spacetime varies, not just in time and space, but across models, and the material content of spacetime influences how it does so.  The fact that the $g_{ab}$ of \textbf{SR2} is the solution of an equation is not a sufficient condition for either of these features.  The $g_{ab}$ of \textbf{SR2} is not affected by matter, because it is wholly determined (up to isomorphism) by equation~\eqref{riem0}.  Relatedly, in the sense that matters, the metric structure of spacetime does not differ from DPM to DPM: the $g_{ab}$s in any two DPMs are isomorphic to one another.\footnote{Strictly, the global topology of the manifold $M$ might allow for infinitely many non-isomorphic flat metric fields.  Even so, these will all be locally isomorphic.}

These features of \textbf{SR2} mean that, if one wishes to remain faithful to the natural pre-theoretic sense of ``background'', it should be classified as a background-\emph{dependent} theory. They further suggest that one should regard \textbf{SR1} and the diffeomorphism-invariant \textbf{SR2} as different formulations of the same, background-dependent theory. In contrast, \textbf{GR1} is (a diffeomorphism-invariant formulation of) a background-independent theory.  This situation might bring to mind Bergmann's claim, noted in Section~\ref{s:dissent}, that the distinctive feature of GR is its lack of a non-generally-covariant formulation.  This feature of GR could not be equated with its background independence:  a background-dependent theory might lack a non-generally-covariant formulation because its background structures lack symmetries.  However, now we have the distinction between general covariance and diffeomorphism invariance on the table, the general approach might appear more promising.

The idea is that it is the \emph{lack} of a \emph{non}-diffeomorphism-invariant formulation, rather than the existence of a diffeomorphism-invariant formulation, that is the mark of a background-independent theory.  A non-diffeomorphism-invariant formulation of a theory requires that some elements of its models are regarded as fixed, identically the same from model to model.  If a theory is background dependent, in the sense that it involves non-dynamical fields that (intuitively) do not vary from model to model, then those fields can be represented by fixed structures in a non-diffeomorphism-invariant formulation of the theory.  But if the theory is background independent, in the sense that all of its fields can vary from model to model, it lacks elements that can be represented by fixed structures.  Of necessity, it will be diffeomorphism invariant.\footnote{This proposal fits with some of the more careful claims from the quantum gravity community concerning the link between background independence and diffeomorphism invariance.  For example, in an informal website article on the meaning of background independence, Baez claims: ``making the metric dynamical instead of a background structure leads to the fact that all diffeomorphisms are gauge symmetries in general relativity'' \citep{baezBKGD}.}  The background fields of a theory are to be identified with those fields that appear as fixed elements in \emph{some} non-diffeomorphism-invariant formulation that theory.  So, for example, the metric field, $g_{ab}$, of \textbf{SR2} represents background structure because it represents the same structure that is represented in the alternative formulation of the theory, \textbf{SR1}, by $\eta_{ab}$.

There is clearly a close connection between identifying a background field in this way and Anderson's notion of an \emph{absolute object} \citep{anderson64,anderson67}.  I will return to this connection at the end of the next section, after reviewing one more complication.

\section{Absolute Objects and the Action--Reaction Principle}\label{s:AO}

Assume that background-independent theories can only be formulated in a diffeomorphism invariant manner.  That leaves open the issue of whether every theory that must be formulated in a diffeomorphism-invariant manner lacks background fields.  Whether one endorses this further claim in part depends on a subtlety concerning what it takes to be a background field.

When the metric field of GR is presented as an example of field that, unlike its precusors in pre-relativistic theories, is not a background field, two of its features are often run together: (i) like other fields in the theory, the metric is \emph{dynamical}; (ii) it also obeys \emph{the action--reaction principle}: it is affected by every field whose evolution it constrains.  The second feature entails the first (assuming the entity in question is not entirely dynamically redundant); a field obviously cannot be dynamically affected and yet not be dynamical.  However, the converse implication does not hold.  A field might affect without being affected and yet have non-trivial dynamics of its own.  

Consider, for example, the theory (call it \textbf{GR2}) given by the following equations:
\begin{gather}
g^{ab}\nabla_a \nabla_b \Phi = 0, \tag{\ref{kggr}} \\
R_{ab} = 0. \label{ricci0}
\end{gather}
Here $R_{ab}$ is the Ricci tensor associated with $g_{ab}$.  In other words, equation~\eqref{ricci0} is the the vacuum Einstein equation, even though the theory's models contain a material scalar field.  In this theory the metric is clearly dynamical; it varies from DPM to DPM.  Since it is constrained to obey equation~\eqref{kggr}, the matter field `feels' the metric.  However, in contrast to the situation in GR, matter does not act back on the metric.  The action--reaction principle is violated.  To adapt Einstein's terminology, as quoted in Section~\ref{s:einstein}, the metric of \textbf{GR2} is a \emph{causal absolute} even though it is a thoroughly dynamical field.

Should $g_{ab}$ count as a background field in this theory?  One might naturally characterise the metric as a background \emph{relative to} the dynamics of $\Phi$.  It is a kind of ``dynamical background field''.  But it does not seem correct to classify \emph{the theory as a whole} as background dependent on this account.  After all, in those models where $\Phi$ vanishes, the theory just is vacuum GR.  This verdict matches that reached if one sticks with the criterion proposed in the previous section (necessary diffeomorphism invariance), for \textbf{GR2} lacks a non-diffeomorphism-invariant formulation in just the way \textbf{GR1} does.

\textbf{GR2} serves another illustrative purpose.  At the end of the previous section I suggested that there is a link between whether a field can appear as a fixed field in a non-diffeomorphism-invariant formulation of a theory and whether that field is an absolute object in Anderson's sense.  Although Anderson informally introduces absolute objects in terms of their violation of the action--reaction principle, the definition he goes on to give characterises them in terms of a notion of sameness in all DPMs of the theory.\footnote{The values of the absolute objects are said to determine the values non-absolute objects but not vice versa (\citealp[83]{anderson67}; \citealp[see also][1658, fn~6]{andersongautreau69}).  In \citet[192]{anderson64}, he says that ``an absolute element in a theory indicates a lack of reciprocity.''  This is consistent with absolute objects being sufficient, but not necessary, for a violation of the action--reaction principle.} What the metric field $g_{ab}$ of \textbf{GR2} illustrates is that a field can be an action--reaction violating causal absolute without being an absolute object in the Andersonian sense.

Let us return to the connection between absolute objects and fixed fields.  How, exactly, are they related?  The answer is not entirely straightforward, partly because different authors define absolute objects slightly differently.

Anderson's formal definition of absolute objects does not characterise them directly.  Instead he defines them in terms of conditions intended to determine when a subset of the dynamical variables of a theory constitute the components of the theory's absolute objects \citep[83]{anderson67}.  \citet[56--60]{friedman83} later advocated a coordinate-free characterisation, according to which a geometric object field counts as absolute if there exist the right kind of maps between any two models of the theory that preserve the object in question (more details shortly).  According to Friedman's set-up, the metric fields of both \textbf{SR1} and \textbf{SR2} count as absolute objects, even though the metric is a fixed field only in \textbf{SR1}.\footnote{Effectively, we are distinguishing two senses of ``dynamical''.  The metric of \textbf{SR2} counts as dynamical in a liberal sense, because it varies non-trivially in the space of KPMs and is constrained to be what it is in any DPM via the ``equation of motion'' \eqref{riem0}.  But in a stricter sense it is not dynamical, because (up to a diffeomorphism) it is the same in every model of the theory.  The stricter sense takes ``dynamical'' to mean ``not absolute''; the liberal sense takes ``dynamical'' to mean ``not fixed''.}  This is not true according to Anderson's definitions.  On his way of setting things up, in a non-covariant coordinate presentation of \textbf{SR1}, there are \emph{no} absolute elements, because the metric field is not explicitly represented \citep[\emph{cf}.][87]{anderson67}.  \emph{In this formulation of the theory}, all of the variables required to characterise a solution (in this case, the values of $\Phi$ relative to some inertial coordinate system) are the components of a genuinely dynamical object.  Nevertheless, it is clear that the metric of \textbf{SR2} counts as an absolute object according to Anderson's definition.  I suggested above that one should regard \textbf{SR1} and \textbf{SR2} as different formulations of the same theory, and thus regard their metric fields as representing the same element of physical reality.  Generalising this move, one can say that an object that features as a fixed field in one formulation of a theory will appear as an absolute object in reformulations of the theory in which that object is no longer treated as fixed.

So far we have noted that fields that are (or can be represented as) fixed are (or can be represented as) absolute objects.  What about the converse?  If a diffeomorphism-invariant theory contains an absolute object, can it be given a non-diffeomorphism-invariant formulation in which that object features as a fixed field?  Here, again, the way Friedman and Anderson define ``absolute object'' makes a difference.  While both, in different ways, formalise a notion of ``sameness in every model'', Anderson's notion of sameness is global whereas Friedman's is local.  More specifically, Friedman holds that, if the models of a theory take the form $\langle M, O_1, \ldots, O_n \rangle$, then object $O_i$ is an absolute object just if, for any two models $\mathcal{M}_1 = \langle M, O_1, \ldots, O_n \rangle$ and $\mathcal{M}_2 = \langle M, O_{1}', \ldots, O'_n \rangle$, and for every $p \in M$, there are neighbourhoods $A$ and $B$ of $p$, and a diffeomorphism $h: A \to B$ such that $O'_i = h^*O_i$ on $A \cap B$.  Friedman's absolute objects can therefore possess ``global degrees of freedom'': differences between such objects might distinguish between classes of DPMs even though the objects are (in the sense just characterised) everywhere locally indistinguishable.\footnote{Consider, for example, flat Lorentzian metrics on a manifold with non-trivial global topology.  Such metrics need not be globally isometric even though they are everywhere flat.  Some models might be temporally finite whereas others are temporally infinite but spatially finite in a preferred spatial direction.}  The upshot is that a theory that involves absolute objects in Friedman's sense may not have a (natural) non-diffeomorphism-invariant formulation in terms of fixed fields.

A popular move is to equate background fields and absolute objects, and so to treat background independence as the lack of absolute objects.  \citet{giulini07} offers a careful recent development of this strategy.  As Giulini notes, and as is discussed in depth by \citet{pitts06}, several ``counterexamples'' suggest that neither Anderson's proposal nor Friedman's get things just right.  The counterexamples come in three categories. (1) There are cases where structure that, intuitively, should count as background is not classified as an absolute.  (2) There are cases where structure that, intuitively, should not count as background is classified as an absolute.  Finally, (3), it is noted that, on Anderson's definition (suitably localised), GR itself turns out to have an absolute object (and so should count as background-dependent).

Torretti's \citeyearpar{torretti84} example of a theory set in classical spacetimes of arbitrary but constant spatial curvature is of type (1).  Pitts observes that if one decomposes the spatial metric into a conformal spatial metric density and a scalar density, then the former is an absolute object while the latter, while constant in space and time, counts as a genuine, global degree of freedom.
  
The best-known case of type (2) is the Jones--Geroch example of the ``dust'' 4-velocity in GR coupled to matter that is characterised by only a 4-velocity field and a mass density.  Pitts sees both Friedman's own suggestion---that one take the 4-momentum field of the dust as primitive \citep[59]{friedman83}---and the option of defining the ``4-velocity'' so that it vanishes in matter-free regions, as motivated by an Andersonian ban on formulations of a theory that contain physically redundant variables \cite[361--2]{pitts06}.\footnote{Pitts pursues the topic further in \citet{pitts09emp}.}  My own view is that both of these ``solutions'' miss the central problem posed by the example.  In the context of this theory, the non-vanishing velocity field is, intuitively, as dynamical as the the 4-momentum.  The trouble arises not because we mistook as indispensable an object that Anderson's definition correctly classifies as absolute.  The trouble is that Anderson's definition, intuitively, misclassifies that object.

The example suggests that the notion of absolute objects might not, in fact, be a better candidate than the notion of fixed fields for articulating the sense of ``dynamical'' relevant to characterising background structure.  Consider, for example, a diffeomorphism-invariant formulation of a theory set in Minkowski spacetime and involving matter characterised, in part, by a (non-vanishing) four-velocity.  One can define two distinct proper subsets of the KPMs (and, correspondingly, the DPMs) of this theory.  The first is obtained by specialising to a particular metric field on the manifold, and retaining all and only those KPMs (and DPMs) that include this metric field.  The second is obtained by specialising to a particular representation of the four-velocity.  If we view each set of models as determining some theory, then both theories involve (in some sense) a fixed field.  However, in the case of the theory obtained by specialising to a particular metric, the solution set is identifiable, as a subspace of the KPMs, via some differential equations for the truly dynamical objects given the fixed field (the metric).  In the case of the ``theory'' with the fixed velocity field, in contrast, it seems highly doubtful that we will be able to view the particular (flat) metrics occurring in the DPMs as all and only the solutions of an equation for the metric \emph{given} the velocity field.  (Imagine specialising to coordinates in which the velocity field takes the value $(1,0,0,0)$ and consider how likely it is that the set of admissible components of the metric field in such coordinates are picked out via an equation.)

A similar strategy might be pursued in the case of (3).  The candidate absolute object in question is the determinant of the metric, $\sqrt{-g}$.  One might accept this verdict without accepting that this automatically means that GR should count as background dependent.  The latter might be held to further require that $\sqrt{-g}$ be interpretable as a fixed field.\footnote{Can the the equations of the theory be interpreted as equations \emph{for} the other variables \emph{given} fixed $\sqrt{-g}$?  This seems to be the correct verdict for unimodular GR, but not (or not clearly so) for GR itself.  For further discussion of this case, although not in terms of the notion of fixed fields, see \citet{earmanCC,pitts06,sus08,susEPSA}.}

Suppose, however, that one sticks with the proposal that the lack of absolute objects is equivalent to background independence.  What light does that shed on the relationship between background independence and diffeomorphism invariance?  Does a theory lack a non-diffeomorphism-invariant formulation just if it lacks absolute objects?  We have seen that, not only are fixed fields not absolute objects (on either Anderson's definition or Friedman's), but \emph{being representable in terms of a fixed field} is also not equivalent to being an absolute object.  Since the presence of fixed fields would seem to be necessary for the failure of diffeomorphism invariance, this means that necessary diffeomorphism invariance cannot be equivalent to background independence understood as lack of absolute objects.

There is a rather desperate way to reconnect the question of whether $\text{Diff}(M)$ is a symmetry group with background independence:  redefine symmetry!  For example, one might try stipulating that $\text{Diff}(M)$ is a symmetry$^*$ group of a theory $T$ iff, if $\langle M, A, D \rangle$ is a model of $T$, then so is $\langle M, A, d^*D \rangle$ for all $d \in \text{Diff}(M)$.  (Formally this looks just the definition of diffeomorphism invariance from Section~\ref{s:gcvsdi}, with ``$F$'', for ``fixed field'' replaced by ``$A$'', for ``absolute object''.)  The proposal is problematic, on at least three grounds.

First, the notion of symmetry$^*$ is transparently ad hoc.  When our theory contained fixed fields, restricting the action of $\text{Diff}(M)$ to the dynamical (i.e., \emph{non-fixed}) fields was natural.  Only by doing so could one define a natural group action on the space of KPMs.  The symmetry group is then naturally defined to be the subgroup of this group that fixes the space of DPMs.  When one has a diffeomorphism-invariant theory that includes absolute objects, one (obviously!)\ does not need to stipulate that $\text{Diff}(M)$ acts only on the dynamical (i.e. \emph{non-absolute}) fields in order for its action on the space of KPMs to be well defined.

Second, defining the action of $\text{Diff}(M)$ on the space of KPMs in such a way that it does not act on the $A$s breaks the natural definition of symmetry.  The definition yields, as intended, that a theory with, say, a flat Lorentzian metric as its absolute object will fail to have $\text{Diff}(M)$ as a symmetry$^*$ group.  But it will \emph{also} fail to have the Poincar\'e group as a symmetry$^*$ group.  For any \emph{given} solution $\langle M, A, D \rangle$, the maximal group $G$ such that, for all $g \in G$, $\langle M, A, g^*D \rangle$ is a solution, will be isomorphic to the Poincar\'e group (or, possibly, a supergroup of the Poincar\'e group).  But for two arbitrary solutions $\langle M, A, D \rangle$ and $\langle M, A', D' \rangle$, the groups so defined need not coincide.  In fact, in general, they will coincide only when $A = A'$.\footnote{Invariance, as I defined it in Section~\ref{s:gcvsdi}, is called \emph{covariance} by Anderson \citeyearpar[75]{anderson67}.  He defines a theory's symmetry, or ``invariance'' group as the ``largest subgroup of the covariance group\ldots which is simultaneously the symmetry group of its absolute objects'' \citep[87]{anderson67}.  It would seem, therefore, that Anderson's symmetry group is related to the notion of symmetry$^*$ in exactly the way the group of automorphisms of the fixed fields of a theory is related the symmetry group (as defined in Section~\ref{s:gcvsdi}) of that theory.  In both cases one should expect the former to be a (possibly proper) subset of the latter.  But we have just seen that, without some finessing, the symmetry$^*$ group of a theory will be trivial.  The same trouble afflicts a flatfooted reading of Anderson's definition.  Consider \textbf{SR2}.  The symmetry group of any \emph{particular} absolute $g_{\mu \nu}$, occurring in a particular DPM, will be (isomorphic to) the Poincar\'e group \citep[\emph{cf}.][87]{anderson67}, but the only diffeomorphism that belongs to every such group is the identity map.}

Suppose one circumvents these problems by adding some epicycles to the definition of symmetry$^*$.  There remains a third reason to be dissatisfied with the proposal that background independence is equivalent to $\text{Diff}(M)$'s being a symmetry$^*$ group.  At bottom, what is doing all the work is the notion of absolute object, in terms of which the gerrymandered notion of symmetry is defined.  If our interest is in characterising background independence, why not simply characterise it as the lack of absolute objects and be done with it?  In particular, the detour via symmetry$^*$ does not give us a better handle on GR's background independence versus SR's background dependence.

\section{$\text{Diff}(M)$ as a Variational Symmetry Group}

When physicists talk of a generally-covariant formulation of a specially relativistic theory, they typically have in mind a formulation like \textbf{SR1}.  Undue focus on such examples, at the expense of examples like \textbf{SR2}, might explain why the connection between background independence and diffeomorphism invariance is sometimes taken to be tighter than it really is.  However, theories along the lines of \textbf{SR2} do get considered by those who defend a diffeomorphism-invariance/background-independence link.  As we have seen, the possibility of such formulations of specially relativistic theories is central to Anderson's thinking (and explains the idiosyncrasies of his definition of symmetry).  The option is also considered by Rovelli, who concedes:
\begin{quotation}
even full diffeomorphism invariance, should probably not be interpreted as a rigid selection principle, capable of selecting physical theories \emph{just by itself}. With sufficient acrobatics, any theory can perhaps be re-expressed in a diffeomorphism invariant language. \ldots

But there are prices to pay. First, [\textbf{SR2}]\ldots has a ``fake'' dynamical field, since $g$ is constrained to a single solution up to gauges, by the second equation of the system. Having no physical degrees of freedom, $g$ is physically a fixed background field, in spite of the trick of declaring it a variable and then constraining the variable to a single solution.  Second, we can insist on a lagrangian formulation of the theory\ldots \citep{sorkin02}, but to do this we must introduce an additional field, and it can then be argued that the resulting theory, having an additional field is different from [the original] \citep{earman89}.  \citep{rovelli07}
\end{quotation}

Several comments are in order.  First, reference to ``sufficient acrobatics'' seems like hyperbole, given the relatively straightforward nature of the transition from a theory like \textbf{SR1} to a reformulation along the lines of \textbf{SR2}.

Second, it is true that, in \textbf{SR2}, $g_{ab}$ is a ``fake'' dynamical field.  It \emph{should} be classified as background structure. Despite our treating it as dynamical in the liberal sense, it remains non-dynamical in a stricter sense.  The previous sections have reviewed apparatus that allows us to draw precisely these distinctions, and to differentiate \textbf{GR1} and \textbf{SR2}, despite both theories being equally diffeomorphism invariant.  So, it is not clear why there is a ``price to pay'' in adopting such a formulation, particularly since we are regarding \textbf{SR2} as merely a reformulation of \textbf{SR1}.  Rovelli, perhaps, would question this last stance.  The diffeomorphism invariance of any theory might be taken to have significant implications for the nature of the true physical magnitudes of the theory, and thus require that one distinguish \textbf{SR2} from (the non-diffeomorphism-invariant) \textbf{SR1}.  If so, I disagree, for reasons I explain in the final section of this paper.

Third, and most interestingly, Rovelli's description of the second cost suggests a quite different way to connect the question of whether diffeomorphisms are symmetries to background independence.  \emph{Prima facie}, there is a formal difference between \textbf{SR2} and \textbf{GR1} that I have not so far mentioned.  The two theories are defined on the same space of KPMs. In the case of \textbf{GR1}, the space of solutions picked out by its equations can also be fixed via a variational problem defined in terms of the action $S_{\text{GR1}} = \int d^4x (\mathcal{L}_G + \mathcal{L}_{\Phi})$.\footnote{The ``gravitational'' part of the Lagrangian is the Einstein--Hilbert Lagrangian $\mathcal{L}_G = \sqrt{-g}\kappa R$, where $R$ is the curvature scalar and $\kappa$ is a suitable constant.   The ``matter'' term is the standard Lagrangian for the massless Klein--Gordon field: $\mathcal{L}_{\Phi} = \sqrt{-g}g^{ab}\nabla_a \Phi \nabla_b \Phi$.}  On the face of it, the same is not true of \textbf{SR2}.  One can pick out the solution space of \textbf{SR1} in terms of a variational problem, defined via the action $S_{\text{SR1}} = \int d^4x \mathcal{L}_{\Phi}$, where $\mathcal{L}_{\Phi}$ depends on the \emph{fixed} metric field $\eta_{ab}$.  In the context of the space of KPMs common to \textbf{GR1} and \textbf{SR2}, however, elements in the solution space of \textbf{SR2} are not stationary points of $\int d^4x \mathcal{L}_{\Phi}$.  The latter can identified by considering the Euler--Lagrange equations one obtains by applying Hamilton's principle to both  $\Phi$ \emph{and} $g_{ab}$.  From the first, one gets the Klein--Gordon equation, but from the second one gets the trivialising condition that the stress-energy tensor for $\Phi$ vanishes.

These reflections might suggest that background independence could be linked to the symmetry status of $\text{Diff}(M)$ in the following way:
\begin{BIV1}
A theory $T$ is background independent if and only if it can be formulated in terms of a variational problem for which $\text{Diff}(M)$ is a variational symmetry group.
\end{BIV1}
Although one can write an action for \textbf{SR1} in a generally-covariant or coordinate-independent manner, $\text{Diff}(M)$ is not a symmetry group of the variational problem that defines the theory's models.\footnote{See \citet[161--2]{belot07} for further discussion of the notion of a variational symmetry.} Recall that the action of $\text{Diff}(M)$ on the \textbf{SR1}'s space of KPMs acts on $\Phi$ but not on $\eta_{ab}$, and does not leave the space of DPMs invariant.  A useful alternative way of stating the proposed condition is as follows:
\begin{BIV2}
A theory $T$ is background independent if and only if its solution space is determined by a generally covariant action \emph{all of whose dependent variables are subject to Hamilton's principle}.
\end{BIV2}
This rules out the generally-covariant version of the \textbf{SR1} action principle, since in this case only $\Phi$ and not $\eta_{ab}$ is subject to Hamilton's principle.  It will also rule out \textbf{SR2} if the solution space of this theory really is not obtainable from an appropriately formulated action principle.

Despite these promising results, the proposal does not work.  In the quotation above, Rovelli refers to \citet{sorkin02}. In that paper, Sorkin, rediscovering a  procedure originally employed by \citet{rosen66}, shows how one can derive equations \eqref{kggr} and \eqref{riem0} from a diffeomorphism-invariant action.  One obtains a Sorkin-type action by replacing $\mathcal{L}_G$ in $S_{\text{GR1}}$ with a different ``gravitational'' term, $\mathcal{L}_S = \sqrt{-g}\Theta^{abcd}R_{abcd}$.  The theory therefore involves a Lagrange multiplier field, $\Theta^{abcd}$, in addition to the fields common to \textbf{SR2} and \textbf{GR1}.  In this new action, all the dependent variables are to be subject to Hamilton's principle.  For ease of reference, let us call the resulting theory (so formulated) \textbf{SR3}. Varying $\Theta^{abcd}$ leads to equation \eqref{riem0}.   Since $\Phi$ does not occur in $\mathcal{L}_S$, varying this field has the same effect as in \textbf{GR1}, and leads to the Klein--Gordon equation \eqref{kggr}.  (One also needs to consider variations of $g_{ab}$.  Rather than the EFE, this leads to an equation that relates $\Theta^{abcd}$, $g_{ab}$ and $\Phi$.)\footnote{Note that the evolution of $\Theta^{abcd}$ is constrained by, but does not affect the evolutions of $g_{ab}$ and $\Phi$.  The action--reaction principle is therefore violated by $\Phi$, with respect to $\Theta^{abcd}$, and not just by $g_{ab}$.  The theory illustrates that requiring that all of the dependent variables in an action be subject to Hamilton's principle does not entail that the resulting theory satisfies the action--reaction principle, \emph{pace} \citet{baezBKGD}.}

Let us assume, for the moment, that in \textbf{SR3} we have yet another way to formulate the specially relativistic theory that has been our example throughout this paper.  Since its models are determined by a diffeomorphism-invariant action, all of whose dependent variables are subject to Hamilton's principle, the theory counts as background independent according to our latest proposal.  The proposal therefore needs to be revised.  A natural thought is to amend it as follows:
\begin{BIV3} A theory $T$ is background independent if and only if its solution space is determined by a generally-covariant action: (i) all of whose dependent variables are subject to Hamilton's principle, and (ii) all of whose dependent variables represent physical fields.
\end{BIV3}
\noindent The idea is that \textbf{SR3} fails to satisfy the second of these conditions because the dynamics of the additional field $\Theta^{abcd}$ strongly suggest that it is not a physical field.  It makes no impact on the evolution of $g_{ab}$ and $\Phi$ and hence, were it a genuine element of reality, it would be completely unobservable (on the natural assumption that our empirical access to it would be through its effect on ``standard'' matter fields such as $\Phi$).  Indeed, it is only on the basis of interpreting $\Theta^{abcd}$ as a mere mathematical device that one can view \textbf{SR3} as a reformulation of \textbf{SR2}.

In the quotation at the start of this section, Rovelli suggests that one might instead regard \textbf{SR3} as a different theory from \textbf{SR2}, on the grounds that \textbf{SR3} involves an additional field (presumably because one views this field as representing a genuine element of reality, the points just made notwithstanding). 
This might seem to provide an alternative way to argue that  our revised proposal does not classify \textbf{SR2} as background independent on the basis of \textbf{SR3}'s satisfying its conditions: if \textbf{SR3} is a different theory, it clearly does not show that the solutions of \textbf{SR2} can be derived from a diffeomorphism-invariant action.

While this might get the classification of \textbf{SR2} correct, it does so at the cost of misclassifying \textbf{SR3}.  According to the current suggestion, \textbf{SR3} now \emph{is} a theory that meets the conditions for being background independent.  But this is not the right result.  The fact the the equation of motion for its metric field is derived from a diffeomorphism-invariant action expressed only in terms of physical fields, hardly makes that metric more dynamical than the metric of \textbf{SR2}.  After all, they both obey exactly the same equation of motion.  And once this problem is recognised, reclassifying $\Theta^{abcd}$ as unphysical does not seem like enough to salvage the proposal.  Even if \textbf{SR3} is no longer a counterexample, might there not be a relevantly similar theory that the proposal incorrectly classifies as background independent? The Rosen--Sorkin method is not the only way to construct a diffeomorphism-invariant variational problem for a theory that involves non-dynamical fields.  These alternative procedures arguably provide examples of exactly the type envisaged.

One such procedure, developed by Karel Kucha\v r, is \emph{parameterization}.  In the simplest case one starts with the \emph{Lorentz-covariant} expression for the action, defined with respect to inertial frame coordinates.  Note that the field $\eta_{ab}$ does not explicitly occur in this expression.  One then treats the four coordinate fields $X^{\mu}$ of this formulation as themselves dependent variables (``clock fields''), writes them as functions of arbitrary coordinates, $X^{\mu} = X^{\mu}(x^{\nu})$, and re-expresses the Lagrangian in terms of these new variables.  Hamilton's principle is applied to the original dynamical variables, now conceived of as functions of $x^{\nu}$, \emph{and} to the coordinate fields, $X^{\mu}$.  In our simple example of \textbf{SR1}, stationarity under variations of $\Phi$ leads to an equation for $\Phi$ and $X^{\mu}$ that is satisfied just if $\Phi$ satisfies the standard Lorentz-covariant Klein--Gordon equation \eqref{kgsr} with respect to the $X^{\mu}$.  Stationarity under variations of the $X^{\mu}$ yields equations that are automatically satisfied if the first equation is satisfied \citep[see, e.g.,][\S II.A]{varadarajanPFT}.  Let us call the resulting theory \textbf{SR4}.

Another technique is described by \citet[734]{leewald90}.\footnote{See \citet[206--9]{belot07} for an extended discussion of this example.}  Let the KPMs of \textbf{SR5} be defined in terms of two maps from the spacetime manifold, $M$.  One is our familiar scalar field $\Phi$.  The other is a diffeomorphism $y$ into a copy of spacetime, $\tilde{M}$, that is equipped with a particular flat Lorentzian metric field.  One can use the diffeomorphism $y$ to pull back the metric on $\tilde{M}$ onto $M$, and use the result, $g_{ab}(y)$, to define the standard Lagrangian, $\mathcal{L}_{\Phi}(y,\Phi) = \sqrt{-g(y)}g(y)^{ab}(\nabla_a \Phi) (\nabla_b \Phi)$, and action functional $S = \int d^4x \mathcal{L}_{\Phi}$.  To determine the theory's solutions we require that $S$ is stationary under variations in both of the theory's fundamental variables, $y$ and $\Phi$.  $\Phi$ variations give us that $\Phi$ satisfies the Klein--Gordon equation with respect to $g_{ab}(y)$.  Variations in $y$ give equations that involve the vanishing of terms that are proportional to $\nabla_n T^n{}_{b}$, where $T^{ab}$ is the stress-energy tensor for $\Phi$.  Since $\nabla_n T^n{}_{b} = \mathbf{0}$ follows from the Klein--Gordon equation, these equations are automatically satisfied. 

Both \textbf{SR4} and \textbf{SR5} are examples of theories defined by diffeomorphism-invariant actions all of whose dependent variables are subject to Hamilton's principle.  They will therefore be counterexamples to our latest proposal just if (i) they are background dependent and (ii) all of their fields are physical fields.  One way to explore whether (i) and (ii) are satisfied is to consider how the theories relate to \textbf{SR2}.  In particular, if they count as reformulations of \textbf{SR2}, then they are formulations of a background dependent theory.

First, recall that a model of \textbf{SR2} is a triple of the form $\langle M, g_{ab}, \Phi \rangle$, where $g_{ab}$ is flat.    A model of \textbf{SR4}, is of the form $\langle M, \Phi, X^0, X^1, X^2, X^3 \rangle$.  That is, it lacks a (primitive) field $g_{ab}$, and includes instead four scalar fields.  Finally, models of \textbf{SR5} are of the form $\langle M, y, \Phi \rangle$, where $y$ is a diffeomorphism into $\tilde{M}$, a copy of $M$ equipped with a fixed metric.

For both \textbf{SR4} and \textbf{SR5}, there is a natural map from that theory's solution space to the solution space of \textbf{SR2}.    For \textbf{SR4}, one first defines the unique flat metric field $g^{X}_{ab}$ associated with the fields $X^{\mu}$ (the metric for which the $X^{\mu}$ are everywhere Riemmann--normal coordinates).  One then requires that the map associates $\langle M, \Phi, X^0, X^1, X^2, X^3 \rangle$ with $\langle M, g_{ab}, \Phi \rangle$ just if $g^{X}_{ab} = g_{ab}$.  For \textbf{SR5}, $\langle M, y, \Phi \rangle$ maps to $\langle M, g_{ab}, \Phi \rangle$ just if $g(y)_{ab} = g_{ab}$.  In the first case, the map is many-one.  The solution space of \textbf{SR4} is intuitively `bigger' than that of \textbf{SR2}.  In the case of \textbf{SR5}, however, the map is a bijection.

This machinery helps articulate how both \textbf{SR4} and \textbf{SR5} can naturally be viewed as reformulations of \textbf{SR2}.\footnote{%
A similar observation can be made concerning \textbf{SR3}.  Its models are of the form $\langle M, g_{ab}, \Phi, \Theta^{abcd} \rangle$ and the map from its solution space to that of \textbf{SR2} simply involves throwing away $\Theta^{abcd}$:  $\langle M, g_{ab}, \Phi, \Theta^{abcd} \rangle \mapsto \langle M, g_{ab}, \Phi \rangle$.  This map is many-one, but the differences between \textbf{SR3} models mapped to the same \textbf{SR2} model concern differences in the non-physical field $\Theta^{abcd}$.
} 
First, consider \textbf{SR4}.  For any model of \textbf{SR2} one can choose special coordinates that encode its metric via the requirement that, in these coordinate systems, $g_{ab} = \text{diag}(-1,1,1,1)$.  In order to understand \textbf{SR4} as a reformulation of \textbf{SR2}, one interprets the fundamental fields of \textbf{SR4} to be such coordinate fields.  So interpreted, \textbf{SR4} is a formulation of a background-dependent theory, since \textbf{SR2} is.  Do the $X^{\mu}$ count as ``physical fields''?  Unlike the $\Theta^{abcd}$ of \textbf{SR3}, they certainly encode something physical, since they encode the metrical facts.  But there is also a sense in which they do not themselves directly represent something physical:  coordinate systems are not physical objects.  Note also that encoding a flat metric via special coordinates in the manner proposed does not uniquely determine the coordinates.  If $\{ X^{\mu}\}$ corresponds to one such set of fields, then so will any set $\{ X'^{\mu}\}$ where the $X'^{\mu}$ are related to the $X^{\mu}$ by a Poincar\'e transformation.  This is the source of the fact that the map from models of \textbf{SR4} to those of \textbf{SR2} is many-one.  This means that (on the suggested interpretation our formalism) the $\{ X^{\mu}\}$ contain some redundancy; ``internal'' Poincar\'e transformations $X^{\mu} \mapsto X'^{\mu}$ should be regarded as mere gauge re-descriptions.

The nature of the bijection between the solution space of \textbf{SR5} and that of \textbf{SR2} makes their interpretation as reformulations of the same background-dependent theory even more straightforward.  Are \textbf{SR5}'s basic variables physical fields?  The dynamical role of $y$ is exhausted by its use to define the pull-back metric on $M$.  It is only through this metric that $y$ enters into the Lagrangian of the theory.  Nonetheless, there is again a clear sense in which the machinery involves arbitrary elements that do not represent the physical facts directly.  In particular, we might have set up the theory in terms of a different (but still flat) metric on the target manifold.  As a mathematical object, this would constitute a different formulation of the theory, and yet the difference does not show up at the level of the pulled-back metrics on $M$: the same range of metrics for $M$ is surveyed, just via different maps to a different object.

The upshot is that it is not clear whether \textbf{SR4} and \textbf{SR5}, interpreted as reformulations of \textbf{SR2}, constitute counterexamples to the proposed criterion for background independence.  All hinges on whether the relevant fields count as physical fields.  They clearly encode physical facts but, equally clearly, they do not do so in the most perspicuous manner.  One might seek to solve this dilemma via further proscriptive modifications to the proposal.  This, of course, risks creating further problems.\footnote{For example, does the metric field of \textbf{GR1} represent the physical facts in the most perspicuous manner?  If \textbf{GR1} is not to count as fully background independent, it should not be on account of this type of failure.}  More importantly, one should recognise that we are now far past the point where one might hope to articulate a simple and illuminating connection between diffeomorphism invariance and background independence.

Rovelli writes:
\begin{quote}
Diffeomorphism invariance is the key property of the mathematical language used to express the key conceptual shift introduced with GR: the world is not formed by a fixed non-dynamical spacetime structure, which defines localization and on which the dynamical fields live. Rather, it is formed solely by dynamical fields in interactions with one another. Localization is only defined, relationally, with respect to the fields themselves. \cite[1312]{rovelli07}
\end{quote}

The moral of our investigation so far is that diffeomorphism invariance cannot be taken to express the shift from non-dynamical to only dynamical spacetime structures. Theories with non-dynamical structure can be formulated in a fully diffeomorphism-invariant manner.  But note that Rovelli's description of the key conceptual shift introduced with GR involves two elements.  In addition to the move from non-dynamical to dynamical spacetime, there is the claim that, in GR, ``localization is only defined, relationally, with respect to the fields themselves.''  I agree that this is how one should understand diffeomorphism-invariant theories.  What the existence of diffeomorphism-invariant formulations of theories with non-dynamical structure indicates, however, is that this feature of a theory is not peculiar to theories that lack non-dynamical fields.  A diffeomorphism-invariant, relational approach to ``localization'' is as appropriate in the context of Newtonian physics and special relativity as it is in GR.  A defence of this claim is the task of the last two sections.

\section{An Aside on the Gauge Status of $\text{Diff}(M)$}\label{s:gauge}

My central claim is this: the observable content of, and the nature of the genuine physical magnitudes of, a specially relativistic theory, whether formulated along the lines of \textbf{SR1} or \textbf{SR2}, are identical in nature to those of an analogue generally relativistic theory, such as \textbf{GR1}.  In the next section I will spell out how this can be so.  In this section, I say a little about when one should interpret diffeomorphisms as gauge transformations.

In the previous section, we saw that Rovelli claimed that \textbf{SR3} might be distinguished from \textbf{SR2} on the grounds that the former involves an additional field.  In the passage quoted above, he cites Earman, who does indeed argue that one should distinguish \textbf{SR3} from more standard formulations of specially relativistic Klein--Gordon theory. Earman's reasoning, however, is rather different from Rovelli's.

\citet{earman06} defines (massive variants of) \textbf{SR1}, \textbf{SR2} and \textbf{SR3}, via the analogues of the equations considered earlier in this paper.\footnote{His equation (3) \citep[451
]{earman06} is (once corrected) the massive analogue of my \eqref{kgsrci}, and defines his \textbf{SR1}-type theory.  His equations (5) and (6) \citep[455]{earman06} are the analogues of \eqref{kggr} and \eqref{riem0}, and define his \textbf{SR2}-type theory.}  (To ease exposition, I use this paper's labels to refer to Earman's theories.)  He is primarily concerned with the comparison between \textbf{SR1} (as obtained from an action principle) and \textbf{SR3}.  Earman's reasons for differentiating the theories, unlike Rovelli's, have nothing directly to do with the presence of an additional field.  He views the theories as distinct because he believes that, in the context of \textbf{SR1}, $\Phi$ can be treated as an observable but, in \textbf{SR3}, it cannot because: (i) only gauge-invariant quantities are observable and (ii) one should regard the $\text{Diff}(M)$ symmetry of \textbf{SR3} as a gauge symmetry.  Earman takes (ii) to be justified by the fact that $\text{Diff}(M)$ is both a local \emph{and a variational} symmetry group in the context of \textbf{SR3}.  In reaching this judgement in this way, he takes himself to be applying a ``uniform method for getting a fix on gauge that applies to any theory in mathematical physics whose equations of motion/field equations are derivable from an action principle'' and that is ``generally accepted in the physics community''  \citep[19]{earmanMCT2}.

As I have argued elsewhere \citep{pooley10}, the fact that this apparatus tells us that $\text{Diff}(M)$ is not a gauge group of \textbf{SR1} is not surprising.  $\text{Diff}(M)$ \emph{is not a symmetry group of} \textbf{SR1} and so \emph{a fortiori} it is not a gauge symmetry group.  What one really wishes to know is whether one should view $\text{Diff}(M)$ as a gauge group of \textbf{SR2}.  Earman does not address this question head-on, but one suspects that his answer would be in the negative, 
for he argues that the solution sets of \textbf{SR1} and \textbf{SR2} are the same \citep[455]{earman06}.  This, of course, simply cannot be correct.  It cannot be the case that (i) $\text{Diff}(M)$ is not a symmetry group of \textbf{SR1}; (ii) $\text{Diff}(M)$ is a symmetry group of \textbf{SR2}; and (iii) the solution sets of \textbf{SR1} and \textbf{SR2} are the same.  It is (iii) that should be given up, and it will be instructive to see where Earman's argument goes wrong.

Here is what he says:
\begin{quote}
The solution sets for [\textbf{SR1}] and for [\textbf{SR2}] are the same, at least on the assumption that the spacetime manifold is $\mathbb{R}^4$. For then there is a global coordinate system $\{x^{\mu}\}$ such that $g_{\mu \nu} = \eta_{\mu \nu}$ (where $\eta_{\mu \nu}$ is the Minkowski matrix) solves [\eqref{riem0}]. Moreover, in this coordinate system [\eqref{kggr}] reduces to [\eqref{kgsrci}\footnote{Since Earman refers to $\eta_{\mu \nu}$ as the Minkowski \emph{matrix}, and since he has switched from Roman indices---which I interpret as signalling coordinate-free, abstract index notation---to Greek indices, it would seem more appropriate to refer to his equation (2), i.e., to equation~\eqref{kgsr}, rather than to his (3).}]. And every solution of [\eqref{riem0}] can be transformed, by a suitable coordinate transformation, into a solution of the form $g_{\mu \nu} = \eta_{\mu \nu}$. Thus, every solution of [\textbf{SR2}] is a solution of [\textbf{SR1}]. Similar reasoning shows that the converse is also true. \citep[455, 466, n~26]{earman06}
\end{quote}
This argument, effectively, ignores the distinction between fields that are solutions to equations and fields that feature in equations as fixed fields.  Here is one way to see the error.  Fix a coordinate system $K$ on $M$ (of the kind Earman considers). Relative to $K$, $\eta_{ab}$ always has the same components in the coordinate representation of every solution of \textbf{SR1}.  Every one of these coordinate descriptions is also a description with respect to $K$ of a solution of \textbf{SR2}.  But, \emph{in addition to these}, every possible set of coordinate functions that one can obtain from the original sets by acting by a diffeomorphism on $\mathbb{R}^4$ also describes---\emph{still relative to $K$}---a solution of \textbf{SR2}.  Note, too, that each of these additional sets of coordinate functions corresponds (relative to $K$) to a representation of a (mathematically, though not necessarily physically) \emph{distinct} solution of \textbf{SR2}.  But these new coordinate functions are not descriptions of solutions of \textbf{SR1} relative to $K$ (the components of the metric tensor have been changed, so they no longer describe $\eta_{ab}$).\footnote{They can be understood as descriptions of solutions of \textbf{SR1}, but only if we allow ourselves to describe things with respect to coordinate systems other than $K$ (in fact, we need to consider one coordinate system for each class related by Poincar\'e transformations).  And when we do this, each solution of \textbf{SR1} is, of course, multiply represented.}

I conclude that Earman's claims do not speak against the natural interpretation of $\text{Diff}(M)$ as a gauge group of \textbf{SR2}.  His own favoured apparatus is simply silent on the question.  When physicists themselves justify the use of the apparatus to identify gauge freedom, they take the deterministic nature of the theories in question as a premise \citep[see, e.g.,][20]{diracLQM}.  In the context of \textbf{SR2}, this premise also leads to the conclusion that $\text{Diff}(M)$ is a gauge group.  In fact, \citet{belotGAUGE} shows how one can regiment the intuitions that are arguably behind such arguments in order to define a notion of gauge equivalence that matches Earman's favoured notion in its verdicts concerning Lagrangian theories but which applies more widely.  Unsuprisingly, Belot's definition tells us that $\text{Diff}(M)$ is a gauge group of \textbf{SR2}.  There remains just one task.  We need to see how this interpretative stance with respect to \textbf{SR2} can be reconciled with an relatively orthodox account of nature of the observables of both background-dependent SR and background-independent GR.

\section{On the Meaning of Coordinates}\label{s:coords}

Recall, again, the similarities between \textbf{GR1} and \textbf{SR2}.  The two theories share a space of KPMs.  They differ only in terms of which subsets of this space are picked out as dynamically possible.  The DPMs of each theory, although distinct sets of mathematical objects, are sets of the same \emph{kind} of objects.  That much is mathematical fact.  These similarities, I submit, make plausible the following interpretative stance: one should treat the two theories uniformly.  On this view, the physical magnitudes of the two theories describe the same types of physical objects.  The theories postulate the same kind of stuff; they just differ over which configurations of this stuff are physically possible.

Why might one reject such a view?  The reason, I think, has to do with a popular, but potentially misleading, way of thinking about the coordinates of non-generally-covariant formulations of pre-relativistic theories.  As I will describe in a moment, this way of thinking about the coordinates of, for example, Lorentz-invariant theories has implications for how one conceives of the content of those theories.  It leads to a way of thinking about the theory's physical content that does not transfer to theories without special coordinates.  The lack of non-dynamical background fields entails (though, as we saw, cannot be equated with) the lack of such coordinates.  It is therefore natural to see the shift from SR to GR, in which background structures are excised, as heralding a radical change in the nature of the content of our physical theories.  Against this, I want to highlight an alternative way of conceiving of the special coordinates of a non-covariant physics. This alternative way is perfectly compatible with the fundamental nature of the content of our physics remaining unchanged in the passage from background dependence to background independence.  It also provides an independently plausible account of the content of background dependent-theories, such as SR.

The influence of the problematic view might well flow from the following passage in Einstein's groundbreaking paper on special relativity:
\begin{quote}
The theory to be developed---like every other electrodynamics---is based upon the kinematics of rigid bodies, since \emph{the assertions of any such theory concern relations between rigid bodies (systems of coordinates), clocks, and electromagnetic processes}. \citep[38, my emphasis]{einstein05sr}
\end{quote}
Einstein seems here to be claiming that the meaning of the theoretical claims of Lorentz-invariant electromagnetism---that is, what those claims are fundamentally about---concerns the relationships between electromagnetic phenomena and rods and clocks.  In other words, the content of the theory's claims is held to be about relationships between electromagnetic phenomena and \emph{material bodies outside of the electromagnetic system under study}.

Versions of this type of view, as an interpretation of the special coordinates of specially-relativistic and Newtonian physics, are explicitly endorsed by, for example, \citet[141--2]{stachel93}, \citet[1592--3]{westmansonego09} and, in several places, Rovelli.  To give a flavour of the importance of the view for Rovelli, I quote at length:
\begin{quotation}
For Newton, the coordinates $\vec{x}$ that enter his main equation
\begin{equation}
\vec{F} = m \frac{\mathrm{d}^2\vec{x}(t)}{\mathrm{d}t^2} \tag{2.152}
\end{equation}
are the coordinates of absolute space.  However, since we cannot directly observe space, the only way we can coordinatize space points is by using physical objects. The coordinates $\vec{x}$\ldots are therefore defined as distances from a chosen system $O$ of objects, which we call a ``reference frame''\ldots

\noindent In other words, the physical content of (2.152) is actually quite subtle:
\begin{description}
\item[]There exist reference objects $O$ with respect to which the motion of any other object $A$ is correctly described by (2.152)\ldots
\end{description}
Notice also that for this construction to work it is important that the objects $O$ forming the reference frame are not affected by the motion of the object $A$.  There shouldn't be any dynamical interaction between $A$ and $O$. \citep[87--8]{rovelli04}\footnote{A similar claim is found in \citet[187--9]{rovelli97}.  There Rovelli combines the claim that in pre-relativistic physics ``reference system objects are not part of the dynamical system studied, their motion\ldots is independent from the dynamics of the system studied'' with the further assertion that the ``mathematical expression'' of the failure of this condition in GR is ``the invariance of Einstein's equations under active diffeomorphisms.''}
\end{quotation}

The similarity with Einstein's claim is clear.  The ``physical content'' of an equation of restricted covariance turns out to involve claims about relations between the dynamical quantities that are explicitly represented in the equations and other material bodies that are only implicitly represented via the special coordinates.  There is one difference worth noting.  For Einstein, the important role of external bodies is to make meaningful spatial and temporal intervals; the bodies in question are rods and clocks. Rovelli, in contrast, emphasises two other roles played by the bodies of his reference system: they fix a particular coordinate system (define its origin) and, more importantly, they define same place over time.  In fact, in spelling out his notion of a material reference system, Rovelli seems to take the notion of spatial distance as primitive and empirically unproblematic.

Now contrast this Einstein--Stachel--Rovelli (ESR) way of understanding special coordinates to what I will call the Anderson--Trautman--Friedman (ATF) perspective (recall footnote~\ref{ATF}), which has already been adopted throughout in this paper.  According to this latter view, a generally-covariant formulation of a theory has the advantage over formulations of limited covariance of making the physical content of the theory fully explicit.  This content includes certain spatiotemporal structures, such as those encoded by the Minkowski metric field $\eta_{ab}$.  In cases where these structures are highly symmetric, one can encode certain physical quantities (e.g., spatiotemporal intervals) via special choices of coordinates adapted to these structures.  Newton's special coordinates are not fundamentally defined in terms of, and Newton's equations do not make implicit reference to, external material bodies.  Rather they are equations that encode physically meaningful chronometric and inertial structure, via certain ``gauge fixing'' coordinate conditions.\footnote{Specifically, one imposes $\Gamma^{\mu}_{\phantom{\mu}\nu \rho} = 0$, $t_{\mu} = (1,0,0,0)$ and $h^{\mu \nu} = \text{diag}(0,1,1,1)$, where $\Gamma^{\mu}_{\phantom{\mu}\nu \rho}$ are the components of the connection, $t_{\mu}$ are the components of the one-form that defines the temporal metric and $h^{\mu \nu}$ are the components of the spatial metric.}

In order to avoid confusion, let me stress that according to both the ESR view and the ATF view the special coordinates of a non-covariant form of pre-relativistic physics have a different meaning to arbitrary coordinates in GR (or a generally covariant form of the pre-relativistic theory).  On both views the special coordinates have physical meaning.  The accounts just differ over what that physical meaning is.

To help further clarify the differences between two views, let me highlight three distinct features that concrete applications of coordinate systems must or may have.
\begin{enumerate}
\item The coordinate system must be anchored to the world in some way.  If it is to be concretely applied, and predictively effective, we must be able to practically determine which coordinate values particular observable events are to be assigned.
\item The coordinate system might be anchored to the world by observable material objects outside of the system under study.  (The system under study might be a proper subsystem of the universe.)
\item The coordinate system might partially encode, or be partially defined in terms of, physically meaningful spatiotemporal quantities (spacetime intervals; inertial trajectories etc.).  In order for this to be applied in concrete cases, we require physical systems that disclose these facts.  Further, these systems may or may not be external to the system being modelled by our theory.
\end{enumerate}

The ATF perspective wholly concerns the third point: the special coordinates of non-generally-covariant formulations of theories encode physical magnitudes.  It is simply silent on the issues raised in the first two points.  The ESR perspective assumes such encoding too, but it makes various further commitments concerning how such coordinate systems are anchored to the world, and what kind of systems disclose the magnitudes that the coordinate systems encoded.  It is important to see that these additional claims are not necessary concomitants of the idea that there is such encoding.

To see this, consider how one might in practice get one's hands on an ATF special coordinate system. The coordinates encode spatial intervals and temporal intervals.  So one needs to be able to measure spatial and temporal intervals.  But without further argument, one's ability to measure these should not be taken to require that the rods and clocks one uses are outside the system that one is describing, much less outside the scope of the theory one is using.  
Note that such spatiotemporal measurement is equally essential to the concrete application of GR, not now to give meaning to special coordinates, but to give empirical content to one of the dynamical fields that is explicitly described.

The ESR idea that, necessarily, special coordinates in pre-relativistic physics gain their meaning from material systems outside the system being studied, blurs the distinction between (i) coordinates encoding physical magnitudes that are disclosed by systems not covered by the theory in question and (ii) the coordinates being anchored to the world via material systems outside the system under study.  Rovelli's idea that ``localisation'' is inherently non-relational in pre-relativistic physics really only relies on (ii).  However, it is easy to see that (ii) is not an intrinsic feature of the special coordinates of pre-relativistic physics.  Even if in practice we often use physical systems to measure spatiotemporal intervals (and thereby fix the ``magnitude-encoding'' aspect of the coordinate system) that we do not (or cannot) actually model in our theory, the anchoring of particular coordinates to the world might simply involve the stipulation that some qualitatively characterisable components of the system under study are to be given such-and-such coordinate values.

Consider the case of a Lorentz-covariant formulation of our theory of the specially-relativistic scalar field, for which $\Phi(x)$ is supposed to be an ``observable'', in contrast to the analogous quantity in GR.  If the special coordinate system in terms of which $\Phi$ is being described is anchored to the world by some reference system not described by the theory, and if the coordinates are understood as encoding objective spatiotemporal quantities, then it is clear what physical meaning $\Phi(x_{0})$ is supposed to have (for any given, particular $x_{0}$) and what the difference in meaning is between the quantities $\Phi(x_{0})$ and $\Phi(x_{0}+\Delta x)$.  However---and this is the absolutely crucial observation---such coordinate representations of $\Phi$ can also be understood to be physically meaningful (in essentially the same way) \emph{without} understanding them in terms of ``non-relational localisation'' thought of as provided by an external anchor for the coordinate system.

Imagine, for example, that one measures $\Phi$ to take a certain value (at one's location).  One stipulates that this value is to be given coordinate values $x_{0}$.\footnote{In reality, in order both to provide a uniquely identifying description of the field that allows us to anchor the coordinate system, and to provide sufficient initial data that a prediction can be extracted from the theory, one should really consider the observation of a certain qualitatively characterisable and spatially extended continuum of field values.  This complication does not alter the basic structure of the story given in the text.}  One then asks what value the theory predicts that the field will take at a certain spatiotemporal distance away from the observed value.  Since such spatiotemporal distances are encoded in the coordinates of the Lorentz-covariant formulation of the theory, this is to ask what the theory predicts the value of $\Phi(x_{0} + \Delta x)$ will be, \emph{given the value of} $\Phi(x_{0})$, where the coordinate difference $\Delta x$ encodes the spatiotemporal interval we are interested in.  Note that, conceived of in this way, $\Phi(x)$ and $\Phi(x+\Delta x)$ specify, not two independently predictable quantities ultimately defined in terms of the relationship of $\Phi$ to an unstated reference object, but a single diffeomorphism-invariant coincidence quantity, involving how the variation of $\Phi$ is related to the underlying metric field $\eta_{ab}$.

If one considers Newtonian physics or special relativity as potentially providing complete cosmological theories, then any anchoring of special coordinate systems has to be done, ultimately, in this second way.  Moreover, any systems that disclose the metric facts are, by hypothesis, describable by the theory.  Of course, this is not how we now understand the empirical applicability of Newtonian physics or special relativity in the actual world.  But the point is that there is no logical incoherence in so conceiving of them.  Indeed, it was the interpretation each was assumed to have prior to 1905 and 1915 respectively.  A theory's including non-dynamical background fields does not, per se, preclude such a cosmological interpretation.

To summarise, the additional commitments of the ESR interpretation of coordinates, over those of the ATF view, are not necessary consequences of a theory's being background-dependent in the sense of involving non-dynamical structure.  The conditions that ESR write into the very meaning of all special coordinate systems might correctly characterise some concrete applications of such systems, but they need not do so.  In fact, sometimes, they do not do so.  Consider, for example, a case whose philosophical importance is stressed by Julian Barbour: the use of Newtonian mechanics by astronomers to determine ephemeris time and the inertial frames.\footnote{For a popular account that stresses the philosophical morals, see \citet[Ch.~6]{barbourEOT}}  Here certain facts about simultaneity and spatial distances are determined ``externally'', but the way the coordinate system is anchored to the world, and the way \emph{some} of the spatiotemporal quantities encoded by the coordinate system are determined (time intervals and an inertial standard of equilocality) are not.

There is, perhaps, one qualification to be made.  I have argued that, in the context of classical background-dependent physics, the ESR story about special coordinate systems does not provide an analysis of their fundamental meaning.  This, however, does not rule out something like the story being correct for background-dependent quantum theory.  In this context, the suggestion would be that certain (non-quantum) background structure in the theory, namely, Minkowski spacetime geometry, really does acquire physical meaning via an implicit appeal to physical systems outside the scope of the theory.  Even if something along these lines were correct (and I register my scepticism), the point to be stressed is that its correctness is not to be understood as flowing from the necessary meaning of such coordinate systems in classical background-dependent physics.

\begin{acknowledgement}
This material began to take something close to its current shape during a period of sabbatical leave spent at UC San Diego.  I thank the members of the UCSD philosophy department for their generous welcome. I have benefitted from correspondence and/or discussions with John Dougherty, Carl Hoefer, Dennis Lehmkuhl, Thomas Moller-Nielsen, Matt Pead, Brian Pitts, Carlo Rovelli, Ad\'{a}n Sus, David Wallace and Chris W\"{u}thrich.  I have also received helpful comments from numerous audience members at talks on related material, in Leeds, Konstanz, Oxford, Les Treilles, Wuppertal, at the Second International Conference on the Ontology of Spacetime in Montreal, at the Southern California Philosophy of Physics group, and at the Laboratoire SPHERE Philosophy of Physics Seminar at Paris Diderot.  Research related to this paper was supported during 2008--10 by a Philip Leverhulme Prize. Finally, special thanks are due to the editor for his encouragement, and to Sam Fletcher and Neil Dewar for numerous comments on an earlier draft.
\end{acknowledgement}


\end{document}